\documentclass[aps,prl,reprint,superscriptaddress]{revtex4-1}

\usepackage{graphicx}

\begin{document}

\title{Hydrodynamics of Diffusion in Lipid Membrane Simulations}

\author{Martin V\"ogele}
\affiliation{Department of Theoretical Biophysics, Max Planck Institute of Biophysics, Max-von-Laue Str. 3, 60438 Frankfurt am Main, Germany}

\author{J\"urgen K\"ofinger}
\affiliation{Department of Theoretical Biophysics, Max Planck Institute of Biophysics, Max-von-Laue Str. 3, 60438 Frankfurt am Main, Germany}

\author{Gerhard Hummer}
\email{gerhard.hummer@biophys.mpg.de}
\affiliation{Department of Theoretical Biophysics, Max Planck Institute of Biophysics, Max-von-Laue Str. 3, 60438 Frankfurt am Main, Germany}
\affiliation{Institute for Biophysics, Goethe University, 60438 Frankfurt am Main, Germany}


\begin{abstract}
By performing molecular dynamics simulations with up to 132 million coarse-grained particles in half-micron sized boxes, we show that hydrodynamics quantitatively explains the finite-size effects on diffusion of lipids, proteins, and carbon nanotubes in membranes. The resulting Oseen correction allows us to extract infinite-system diffusion coefficients and membrane surface viscosities from membrane simulations despite the logarithmic divergence of apparent diffusivities with increasing box width. The hydrodynamic theory of diffusion applies also to membranes with asymmetric leaflets and embedded proteins, and to a complex plasma-membrane mimetic.
\end{abstract}

\maketitle

Molecular dynamics (MD) simulations provide insight into the organization and dynamics of lipids and membrane proteins \cite{pluhackova2015a,ingolfsson2016a,hedger2016a,duncan2017a}. Receptor clustering, lipid second-messenger patterning, and lipid domain formation occur in systems with complex lipid composition \cite{ingolfsson2014a} on length scales $\ge$100 nm. Advances in computing, coarse-grained force fields \cite{marrink2007a,cooke2005a,izvekov2005a}, and simulation management \cite{wassenaar2015b,stansfeld2015a,wu2014a} open up this biologically important regime to simulations \cite{ingolfsson2016a,koldso2016a,lyman2018a}. However, simulations of dynamics in membranes face a serious challenge: the translational diffusion coefficients of membrane-embedded molecules are ill-defined. As anticipated from hydrodynamic theory \cite{camley2015a} and shown by MD simulations \cite{voegele2016a,venable2017a}, the apparent diffusion coefficients diverge logarithmically with the size of the simulated membrane patch.
One can think of a membrane particle and its periodic images above and below as forming an infinite quasi-cylindrical structure embedded in a layered medium that effectively imposes 2D flows. In this picture, the logarithmic divergence of the diffusion coefficient is a molecular-scale manifestation of Stokes' paradox, i.e., the vanishing hydrodynamic friction of an infinite cylinder in an infinite medium with 2D flow.
The divergence appears to preclude a meaningful comparison between simulation and experiment for membrane dynamic processes.

Here we show that hydrodynamic theory \cite{camley2015a,voegele2016a} can be used to overcome this challenge, as in neat fluids \cite{yeh2004b}. First, we show that the logarithmic divergence can be broken by expanding the system also in the third dimension, normal to the membrane. This requires simulations with $\ge$10$^8$ coarse-grained particles. Then we show that the Oseen correction, a hydrodynamic correction using the Oseen tensor for a point perturbation \cite{voegele2016a}, quantitatively accounts for the observed behavior, from lipids to membrane proteins and over the entire range of box widths and heights. On this basis, we develop a procedure to correct the simulated diffusion coefficient. By exploiting the strong finite-size dependence, we not only extract the true infinite-system diffusion coefficients $D_0$ of lipids or embedded proteins, but also the difficult-to-obtain membrane surface viscosity $\eta_m$. We apply the formalism to simulations of the diffusion of proteins embedded in lipid membranes, and of a plasma-membrane model with a complex lipid composition.

For neat \cite{Dunweg1993,yeh2004b} and confined fluids \cite{simonnin2017a}, hydrodynamic self-interactions under periodic boundary conditions (PBC) account for the system-size dependence of self-diffusion coefficients ${D}_\mathrm{PBC}$ in MD simulations,
\begin{equation}
  \label{eq:1}
  {D}_\mathrm{PBC} = {D}_0 
  + k_{\mathrm{B}}T \lim_{r\to 0}
  \mathrm{Tr}[\mathbf{T}^\mathrm{PBC}(\vec{r}) 
  - \mathbf{T}_0(\vec{r})]/n_d
\end{equation}
In the Oseen correction, $\Delta D=D_\mathrm{PBC}-D_0$ is approximated as the difference between the Oseen tensors $\mathbf{T}^\mathrm{PBC}(\vec{r})$ for PBC and $\mathbf{T}_0(\vec{r})$ for the infinite system at the origin, $r\to 0$, with $\mathrm{Tr}$ the trace, $n_d$ the dimension ($n_d=2$ for membranes), $k_{\mathrm{B}}$ the Boltzmann constant, and $T$ the absolute temperature.

This formulation suggests hydrodynamic corrections also for membrane simulations \cite{camley2015a,voegele2016a}.
In the Saffman-Delbr{\"u}ck (SD) model \cite{saffman1975a,hughes1981a,petrov2008a}, the membrane is treated as a viscous fluid embedded in an infinite solvent system. Camley et al. \cite{camley2015a} extended the SD model to PBC by representing the Oseen tensor as a two-dimensional lattice sum, 
$T^{\mathrm{PBC}}_{ij}(\vec{r}) = L_x^{-1}L_y^{-1}\sum_{\vec{k}\neq 0} \tau_{ij}(\vec k) \, \exp( -i \vec k \cdot \vec r )$, where 
$\tau_{ij}(\vec{k}) = ( \delta_{ij} - {k_i k_j}/{k^2} )/(\eta_m k^2 + 2 \eta_f k \tanh(k H))$. 
The ratio of membrane-surface and solvent viscosities $\eta_m$ and $\eta_f$,  respectively,  defines the SD length $L_\mathrm{SD} = \eta_m/2\eta_f$. The wave vectors are $\vec{k}=2\pi(n_x/L_x,n_y/L_y)$ with $n_i$ integers and $L_i$ the box widths ($i=x,y$), $k=|\vec{k}|$, and $\delta_{ij}$ the Kronecker delta. $2H=L_z-h$ is the height of the solvent layer separating the periodic images of the membrane, with $h$ the membrane thickness and $L_z$ the box height. The $\tanh$ term accounts for the influence of the surrounding solvent on the diffusion inside the membrane. An Oseen tensor for monotopic inclusions (spanning only one leaflet, such as typical lipids) was proposed as \cite{camley2015a}:  $\tau_{ij}(\vec{k}) = ( \delta_{ij} - {k_i k_j}/{k^2} ) {A(k)}/({A(k)^2-B(k)^2})$, with $A(k) = \eta_m k^2 /2 + \eta_f k/\tanh(2Hk) + b$, $B(k) = \eta_f k/\sinh(2Hk) + b$, and $b$ the interleaflet friction coefficient. We sped up convergence of the lattice sums in Eq.~(\ref{eq:1}) by adding and subtracting integrals \cite{voegele2016a} that can be solved analytically for the transmembrane case and numerically for the monotopic case (see Supplemental Material \cite{SI}). All correction formulas are implemented in Python and available at \url{https://github.com/bio-phys/memdiff} along with an example application.

For the diffusion in membranes contained in flat square simulation boxes, $L=L_x=L_y\gg L_z$, one has \cite{voegele2016a} 
\begin{equation}
\label{eq:2}
D_\mathrm{PBC} \approx
   D_0 + \frac{k_{\mathrm{B}} T}{4\pi\eta_m} \frac{\ln\left( L/( L_\mathrm{SD}  + 1.565 H )\right) - 1.713}{\left(1 + H/L_\mathrm{SD}\right)}
\end{equation}
Accordingly, $D_{\mathrm{PBC}}$ diverges asymptotically as $\ln L$ for large widths $L$ and fixed height $H$ (or $L_z$).  This approximation is also valid for narrow boxes, $L < L_\mathrm{SD}$, if one sets $H=0$ instead of using the actual value \cite{voegele2016a}.  At a box width of $L_c \approx \left( L_\mathrm{SD} + 1.565 H \right)\, e^{1.713}$, in-plane and between-membrane self-interactions effectively cancel, and the box-size corrections vanishes, $D_\mathrm{PBC} \approx D_0$. In numerical tests, the flat-box approximation Eq.~(\ref{eq:2}) is within 2 {\%} of Eq.~(\ref{eq:1}) for atomistic and coarse-grained systems (see Supplemental Material \cite{SI}). The hydrodynamic correction (but not $D_0$!) is insensitive to variations in the interleaflet friction coefficient $b$ for typical lipid models (see Supplemental Material \cite{SI}). The simpler transmembrane correction is thus expected to be an excellent approximation also for monotopic molecules such as individual lipids.

\begin{figure}[tb]	 \includegraphics[width=\columnwidth]{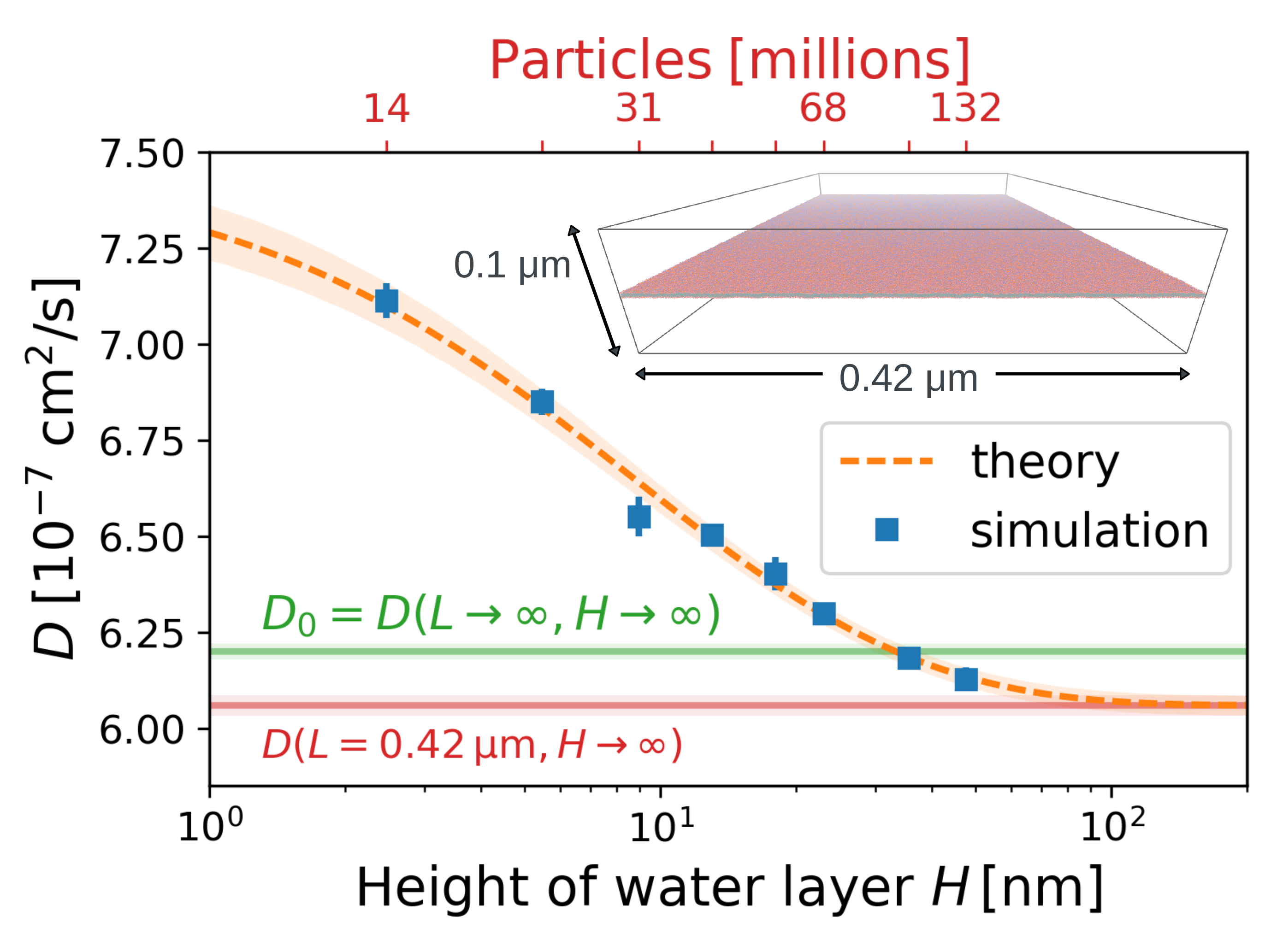} 
  \caption{ \label{fig:figure1} Diffusion coefficients of POPC lipids from MD simulations (symbols) as function of the height $H$ of the water layers above and below the membrane in simulation boxes of constant width $L = 0.42\,\mu$m (top axis: number of particles in the system). The prediction according to Eq.~(\ref{eq:1}) with the transmembrane Oseen tensor is shown as orange dashed line. $D_0$ and $\eta_m$ were fitted to POPC membrane simulations in flat boxes \cite{voegele2016a} and $\eta_f$ was determined independently from pressure fluctuations in bulk water simulations \citep{SI}. 
Horizontal lines indicate the true diffusion coefficient $D_0$ (green) and the limit $H\to\infty$ for fixed $L$ (red).
Shading indicates the uncertainty range (1 s.d.). 
}
\end{figure}

Key open questions are: (1) does the Oseen correction apply beyond the flat-box limit with its logarithmically divergent $D_{\mathrm{PBC}}$; (2) how can one extract meaningful diffusion coefficients from membrane simulations; and (3) does a simple materials parameter, $\eta_m$, suffice to describe the dynamics in complex asymmetric membranes?  
To address the first challenge, we performed simulations with boxes large also normal to the membrane, $L_z\gg L_\mathrm{SD}$. To be consistent with \cite{voegele2016a}, we simulated lipid membranes using the MARTINI coarse-graining scheme \cite{marrink2007a} and the GROMACS 4.5.6. software package \cite{hess2008a}. The bilayer structures were built using \texttt{insane.py} \cite{wassenaar2015b}. Water was added to reach the desired box heights. Because undulations of the lipid bilayer shorten the distance of lipid motions projected onto the $x$-$y$ plane, we suppressed long-wavelength undulations by a weak harmonic restraint acting on the $z$ coordinate of the center of mass of a quarter of the lipids \cite{ingolfsson2014a,voegele2016a}. With these restraints, we assure a constant wavelength spectrum of undulations over all box sizes. Otherwise, long-wavelength undulations would only be suppressed in small boxes, with the longest wavelengths  permitted under PBC being $L_x$ and $L_y$. Energy minimization was followed by equilibration and data production runs in an NPT \cite{berendsen1984a,bussi2007a} ensemble with semiisotropic pressure coupling at 1 bar and 300 K.  Simulation details are listed in the Supplemental Material \cite{SI}.

We obtained diffusion coefficients and viscosities by minimizing $\chi^2=\sum_{i=1}^N (D_i-D_\mathrm{PBC}^{(i)})^2/\sigma_i^2$ with respect to $D_0$ and $\eta_m$, treating $\eta_f$ either as an additional parameter in the minimization or fixing it at the bulk water viscosity, as determined from independent simulations. $i$ indexes the $N$ runs with different box sizes. $D_i$ is the uncorrected diffusion coefficient of run $i$ and $D_\mathrm{PBC}^{(i)}=D_0+\Delta D^{(i)}$ with $\Delta D^{(i)}$ the Oseen finite-size correction Eq.~(\ref{eq:1}) evaluated numerically for fixed membrane thickness $h=4.5$~nm as described in Supplemental Material~\cite{SI}.

The $D_i$ were determined from the slopes of straight-line fits to the mean-squared displacement (MSD) in the membrane plane \cite{voegele2016a} over a time window from 40-90 ns for lipids, 4-9 ns for membrane-spanning carbon nanotubes (CNT; see \cite{voegele2016a,voegele2018a} for details on the CNT model), and 20-40 ns for integral membrane proteins, using shorter times for the latter two because their low abundance affects the sampling at longer times. We calculated the MSD with a Fourier-based algorithm \cite{calandrini2011a}, after removing the center-of-mass motion of the membrane from lipid, protein, and CNT trajectories. Statistical errors $\sigma_i$ were estimated by block averaging using 20 blocks.

Figure \ref{fig:figure1} shows that Eq.~(\ref{eq:1}) accounts quantitatively for the calculated diffusion coefficients for systems with up to 132 million particles in simulation boxes $L=0.42\,\mu$m wide and up to $L_z=0.1\,\mu$m tall. The simulation results match the hydrodynamic predictions using $\eta_m$ fitted only to flat-box simulations \cite{voegele2016a} and $\eta_f=10.2(4)$ determined independently from pressure fluctuations \cite{hess2002a} of bulk MARTINI water. From a global fit of the transmembrane Oseen correction against all POPC simulations here and in \cite{voegele2016a}, we obtained $D_0 = 6.20(2)\times 10^{-7}$ cm$^2$/s, $\eta_f = 9.6(2)\times 10^{-4}$ Pa~s, and $\eta_m = 3.97(6)\times 10^{-11}$ Pa~s~m, so that $L_\mathrm{SD}=20.7$ nm. The monotopic correction with $b=2.9\times 10^6$ Pa~s/m \cite{denotter2007a,SI} gives an indistinguishable fit with the same $D_0$, $\eta_f = 10.17(20)\times 10^{-4}$ Pa~s, and $\eta_m = 4.05(6)\times 10^{-11}$ Pa~s~m. Thicker water layers weaken between-membrane hydrodynamic interactions and slow down lipid diffusion. For the tallest systems, $D_\mathrm{PBC}$ approaches a plateau. However, even with $10^8$ particles, the turnover is incomplete. The limit for $H\to\infty$ is below $D_0$; i.e., for tall boxes, hydrodynamics retards diffusion.

\begin{figure}[tb]
 	 \includegraphics[width=\columnwidth]{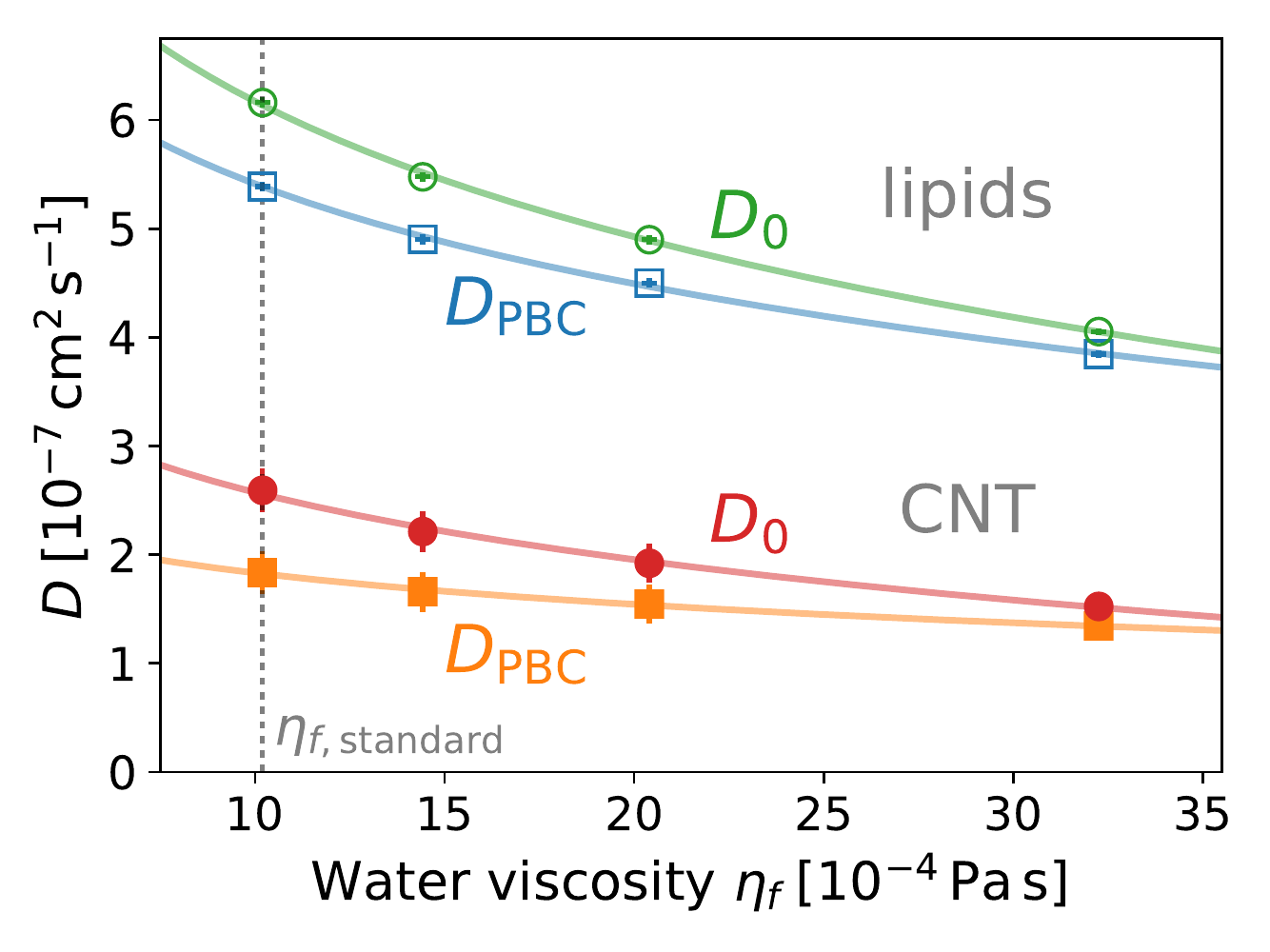} 
         \caption{\label{fig:figure2} Dependence of diffusion coefficients $D_{\mathrm{PBC}}$ and $D_0$ of POPC lipids and CNTs \cite{voegele2016a} on water viscosity $\eta_f$ ($L=40$ nm, $L_z=9$ nm, 300 K). Lines show theory (Eq.~(\ref{eq:1}) and $D_0^{\mathrm{SD}}$). Vertical line: $\eta_f$ for standard MARTINI water. }
\end{figure}

As an additional test of the hydrodynamic model, we examined the effect of water viscosity on diffusion in the membrane (Fig. \ref{fig:figure2}). We reduced the SD length $L_{\mathrm{SD}}$ by increasing the mass $M$ of MARTINI water particles up to 10-fold, scaling the water viscosity as $M^{1/2}$ without altering the structure and thermodynamics of the system. For large $\eta_f$, $L_\mathrm{SD}$ becomes small and $D_\mathrm{PBC}$ approaches the infinite-box limit $D_0$. As shown in Fig. \ref{fig:figure2}, the water viscosity dependence of the diffusion coefficients both of lipids in a neat membrane and of membrane-spanning CNTs quantitatively agrees with the predictions of Eq.~(\ref{eq:1}), further validating the hydrodynamic model.


We determined hydrodynamic radii $R_h$ of diffusing molecules by setting their $D_0$ equal to the SD expression for the diffusion coefficient \cite{saffman1975a}, $D_0^{\mathrm{SD}}=k_{\mathrm{B}}T(4\pi\eta_m)^{-1}(\ln(\eta_m/\eta_fR_h)-\gamma)$ with $\gamma\approx 0.5772$ the Euler-Mascheroni constant. For the CNT, a fit to the data in \cite{voegele2016a} gave $D_0 = 2.76(12)\times 10^{-7}$ cm$^2$/s. The resulting hydrodynamic radius of $R_h = 0.83(14)$ nm agrees with the geometric value of 0.85 nm for this ideal cylinder obtained by summing the radii of the cylinder (0.615 nm) and a carbon bead (0.235 nm). Values of $D_\mathrm{PBC}\le 1.9\times 10^{-7}$ cm$^2$/s without hydrodynamic correction would have given unphysical radii $R_h\ge 2.3$ nm.

 \begin{figure}[bth]
  \includegraphics[width=\columnwidth]{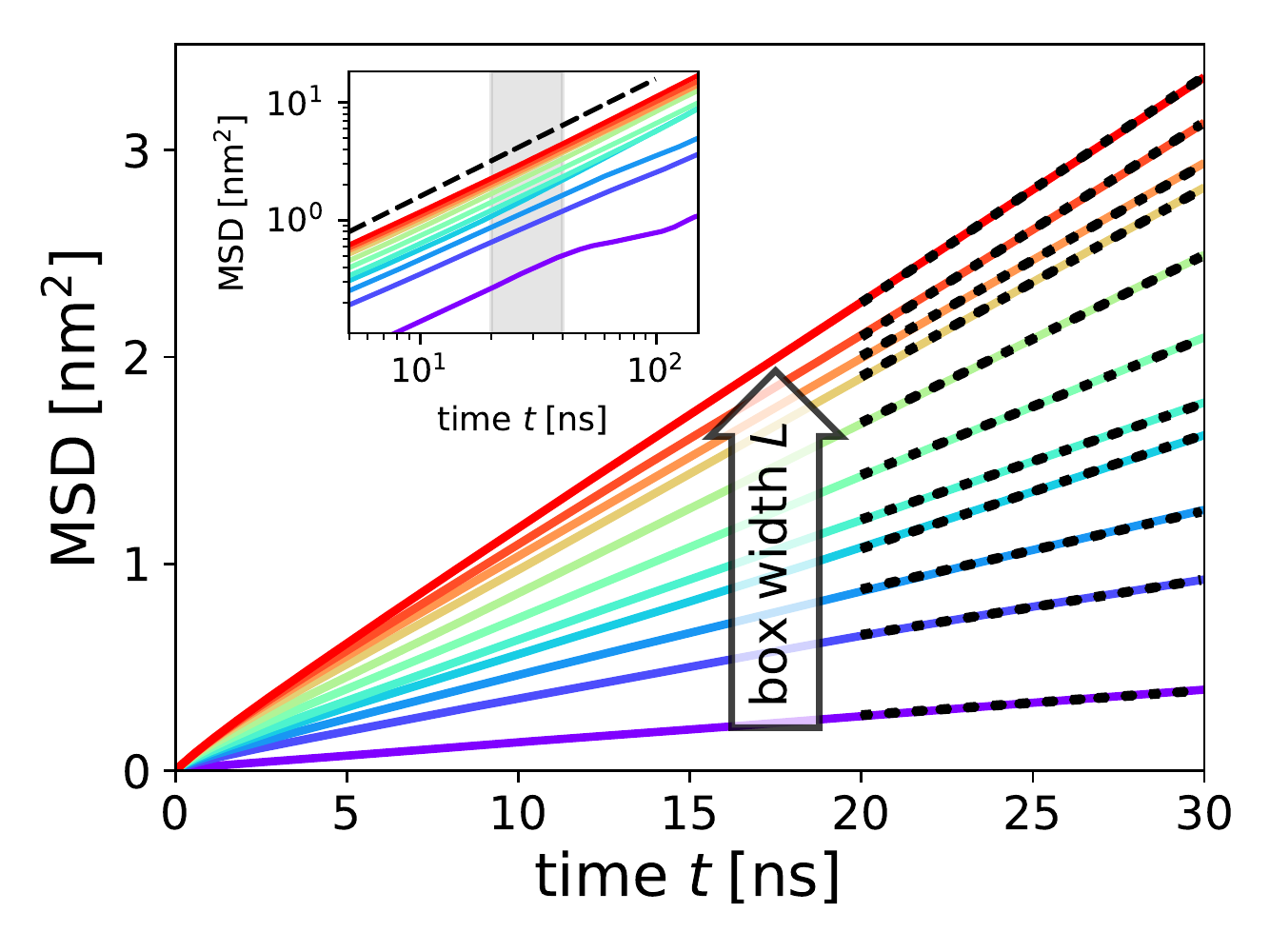}
  \caption{MSD of protein ANT1 in model mitochondrial membrane simulated in boxes of widths from $L=12$ nm (purple) to 360 nm (red) at constant box height $L_z=10.2$ nm ($H=2.85$ nm). Dotted black lines show fits of $\mathrm{MSD}(t) = a + 4 D t$ over the time window used to extract the uncorrected diffusion coefficients $D$. The intercept $a$ accounts for local and fast molecular motions before proper diffusion sets in. In the double-logarithmic inset, the dashed line indicates a linear dependence on time. The fitting region is highlighted in gray.
  \label{fig:figure3}}
\end{figure}

\begin{figure*}[tbh]
	\includegraphics[width=\textwidth]{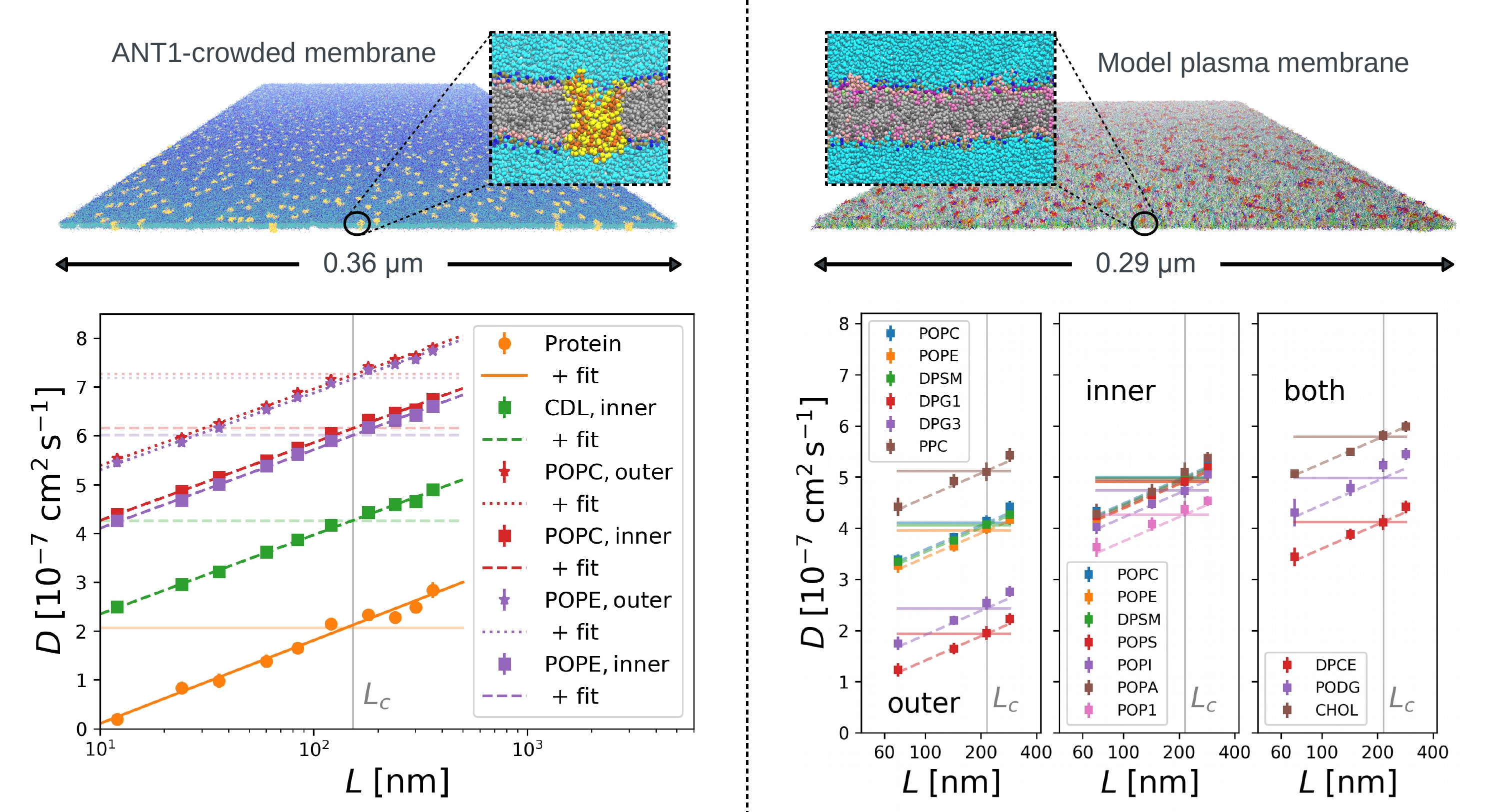}
    \caption{ \label{fig:figure4}
      Diffusion in complex membranes. Upper left: Model inner mitochondrial membrane ($L=360$ nm) with 900 ANT1 transmembrane proteins (yellow; see zoom-in). Lower left: $D_{\mathrm{PBC}}$ (symbols) and fits to hydrodynamic theory (protein: circles and solid lines; outer leaflet: stars and dotted lines; inner leaflet: squares and dashed lines), and infinite-system values (horizontal lines). Upper right: Plasma-membrane simulation ($L=286$ nm). Lower right: $D_{\mathrm{PBC}}$ for representative membrane components that remain in each leaflet (inner/outer) and for those that jump between both leaflets (``both''). Vertical gray lines indicate $L=L_c$, where $D_\mathrm{PBC}=D_0$. Corrections here were calculated using Eq.~(\ref{eq:2}).
}
\end{figure*}

Hydrodynamic theory also accounts for the dramatic box-size dependence of membrane protein diffusion (Fig.~\ref{fig:figure3}). As a model inner mitochondrial membrane, we simulated a POPC/POPE membrane densely packed with the membrane-spanning protein adenine nucleotide translocase (ANT1) and with cardiolipin in the inner leaflet \cite{hedger2016a}. Systems were built with MemProtMD \cite{stansfeld2015a} for a wide range of box widths $L$ (Fig. \ref{fig:figure4} upper left) at fixed heights $H$, such that proteins covered $\approx$11 {\%} of the membrane area while not yet forming large clusters within the simulation time. In simulations using the parameters of \cite{hedger2016a}, the proteins and different lipid components exhibit the same finite-size dependence  (Fig. \ref{fig:figure4} lower left). Despite variations by about a factor 50 for the smallest systems, the apparent diffusivities $D_\mathrm{PBC}$ grow linearly as function of $\ln L$ with component-independent slopes, as predicted by the Oseen correction.  Diffusion of POPC and POPE is slower in the inner leaflet by about 20 {\%}, likely due to the presence of the large cardiolipin molecules.  The ANT1 mitochondrial model membrane has an effective viscosity $\eta_m\approx 4.36 \times 10^{-11}$ Pa~s~m, and ANT1 has a hydrodynamic radius of $R_h=2.1(4)$ nm, close to $R_h=2.3$ nm estimated from the convex hull in the $xy$ plane. By contrast, uncorrected diffusion coefficients would have given  $R_h$ from 0.7 to 24.3 nm.

The Oseen correction also applies to membranes of even more complex composition. Figure \ref{fig:figure4} (right) shows that finite system sizes affect the diffusion in a plasma-membrane model \cite{ingolfsson2014a}. We used the simulation parameters and the configuration provided at \url{http://cgmartini.nl} and built start configurations as squares of 1, 4, 9, and 16 copies of the original box. Even without clear phase separation \cite{ingolfsson2014a},  heterogeneous structures emerged as small clusters of lipids. Moreover, molecules such as cholesterol flipped between the leaflets. Nevertheless, 
the slope of the apparent diffusion coefficients $D_{\mathrm{PBC}}$ with respect to $\ln L$ is independent of membrane component and leaflet localization, defining an effective membrane viscosity $\eta_m \approx 4.73\times 10^{-11}$ Pa~s~m according to Eq.~(\ref{eq:2}) (Fig.~\ref{fig:figure4} lower right). Even in asymmetric membranes of complex composition, a component-independent correction compensates for large finite-size effects.

We showed that finite-size effects in membrane simulations can be corrected by hydrodynamic theory. The Oseen corrections are independent of membrane component. Complex lipid composition and integral membrane proteins do not alter the effects in absence of protein clustering \cite{duncan2017a}, strong protein crowding \cite{javanainen2017a}, and phase segregation. With the Oseen correction Eq.~(\ref{eq:1}) and its approximation Eq.~(\ref{eq:2}), two simulations in flat boxes of different widths $L$ suffice to determine proper membrane diffusion coefficients $D_0$ and  membrane viscosities $\eta_m$, using $\eta_f$ from independent bulk-solvent simulations. 
Thermostats are used in standard protocols for membrane MD simulations. Nevertheless, for weakly coupled rescaling thermostats \cite{berendsen1984a,bussi2007a}, the diffusion of lipids, proteins, and nanotubes in membranes follows the predictions of hydrodynamic theory with respect to the dependence on system size and water viscosity. Based on the remarkable accuracy in capturing the dynamics of complex lipid membranes, we expect the hydrodynamic model to apply to transport phenomena also in other 2D layered materials.

\section{Acknowledgments}

We thank Frank L. H. Brown, Richard W. Pastor, Lukas S. Stelzl, and Max Linke for helpful discussions. We acknowledge PRACE for access to \textit{Mare Nostrum} at the Barcelona Supercomputing Centre. Further computations were performed on \textit{Hydra} at the Max Planck Computing and Data Facility Garching. This work was supported by the Max Planck Society.


%

\onecolumngrid
\clearpage
\begin{center}
\textbf{\large Supplemental Material:\\[1ex] Hydrodynamics of Diffusion in Lipid Membrane Simulations}\\[0.5\baselineskip]
Martin V\"ogele,$^1$ J\"urgen K\"ofinger,$^1$ and Gerhard Hummer$^{1,2}$\\[0.5\baselineskip]
$^1$\textit{Department of Theoretical Biophysics, Max Planck Institute of Biophysics, Frankfurt am Main, Germany}\\
$^2$\textit{Institute for Biophysics, Goethe University, Frankfurt am Main, Germany}
\end{center}
\setcounter{equation}{0}
\setcounter{figure}{0}
\setcounter{table}{0}
\setcounter{page}{1}
\setcounter{section}{0}
\makeatletter
\renewcommand{\theequation}{S\arabic{equation}}
\renewcommand{\thefigure}{S\arabic{figure}}
\renewcommand{\thetable}{S\arabic{table}}


\vspace{8ex}

\section{Theoretical Details}

\begin{figure}[ht]
 \includegraphics[width=1.0\linewidth]{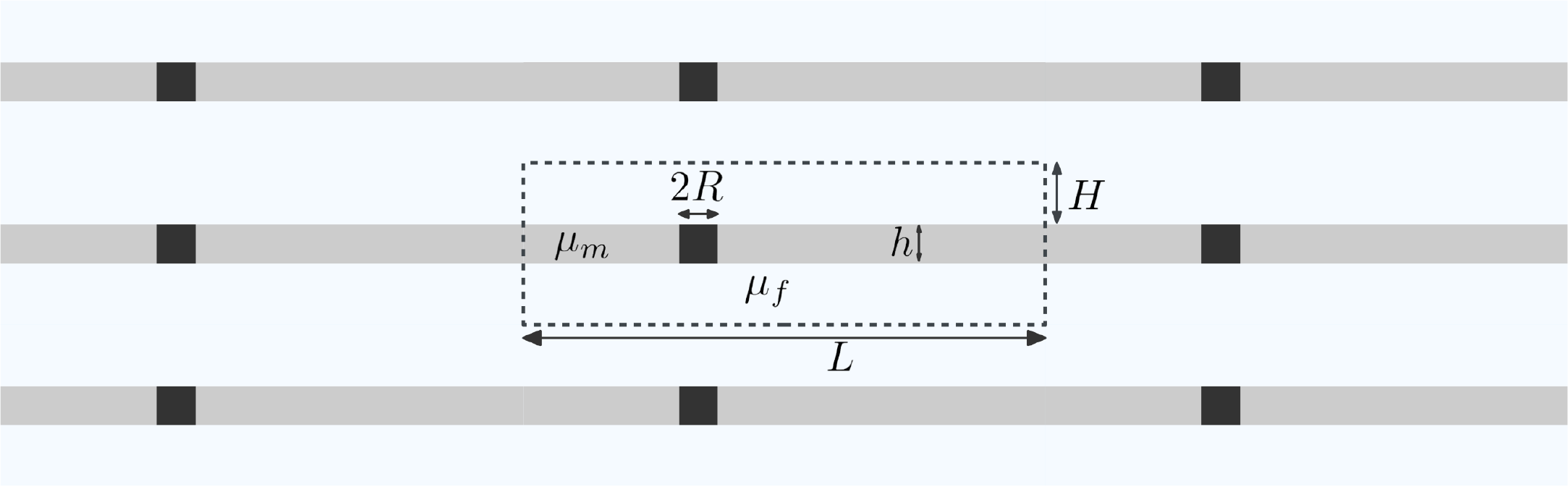}  
 \caption{Cut through a schematic simulation box (dashed lines) of a lipid membrane (gray) with surrounding periodic images according to the Saffman-Delbr{\"u}ck model. The membrane itself is treated as a fluid layer of height $h$ and viscosity $\eta_m$ immersed in a fluid (light blue) of viscosity $\eta_f$, with a cylindrical inclusion (black) of radius $R$. }
  \label{fig:membrane_periodic}
\end{figure}

\subsection{Numerical Evaluation of the Correction Formulas for Transmembrane and Monotopic Inclusions}

The numerical evaluation of the correction formula for transmembrane inclusions was described in \cite{voegele2016a}. We subtract the long-wavelength part from the lattice sum and approximate it by an integral that can be solved analytically:
\begin{eqnarray}
  \label{eq:dT_transmembrane}
  2\Delta T ^\mathrm{trans}  &\equiv & \frac{1}{L^2}\sum_{\vec{k}(\ne 0)}
  \frac{1} {\eta_m k^2 + 2 \eta_f k
      \tanh(k H)} 
      - \int \frac{d^2k}{(2\pi)^2} \frac{1} {\eta_m k^2 + 2
                                \eta_f k} \\
 \label{eq:dT_transmembrane_solution}
 & = & \lim_{\sigma\to\infty}\Bigg[ \frac{1}{L^2}\sum_{\vec{k}(\ne 0)}
  \Bigg(\frac{1} {\eta_m k^2 + 2 \eta_f k
      \tanh(k H)}  
      -\frac{1-e^{-k^2/2\sigma^2}} {\eta_m k^2 + 2 \eta_f k}\Bigg)
 -\frac{e^{-u^2}\left[
\pi\,\mathrm{erfi}(u)-E_i(u^2)\right]}{4\pi\eta_m}\Bigg],
\end{eqnarray}
where $\Delta T = \lim_{r\to 0} \mathrm{Tr}[\mathbf{T}_\mathrm{PBC}(\vec{r}) - \mathbf{T}_0(\vec{r})]/2$, $u=\sqrt{2}\eta_f/\eta_m\sigma$, $\mathrm{erfi}(x)$ is the imaginary error function, and $E_i(x)$ the exponential integral. 

In the monotopic case (an inclusion only within one leaflet or a lipid), we follow the same strategy, with the difference that the integral is solved numerically instead of analytically. Again, we first perform the subtraction of the long-wavelength part:
\begin{eqnarray}
\label{eq:dT_monotopic}
2 \Delta T ^\mathrm{mono}  &\equiv & \frac{1}{L^2} \sum_{\vec{k}(\neq 0)} \frac{A(k)}{A(k)^2-B(k)^2} - \int \frac{\mathrm{d}^2 k}{(2\pi)^2} \left[\frac{A(k)}{A(k)^2-B(k)^2} \right]_{H\rightarrow\infty}\\[2ex]
&=& \frac{1}{L^2} \sum_{\vec{k}(\neq 0)} \left[ \frac{A(k)}{A(k)^2-B(k)^2} - \frac{\eta_\mathrm{mono} k^2 + \eta_f k + b }{(\eta_\mathrm{mono} k^2 + \eta_f k + b)^2 - b^2} \right] \nonumber\\
&+& \lim_{\sigma\to\infty}\Bigg[ \frac{1}{L^2}  \sum_{\vec{k}(\neq 0)}  \frac{e^{-k^2/2\sigma^2} (\eta_\mathrm{mono} k^2 + \eta_f k + b)}{(\eta_\mathrm{mono} k^2 + \eta_f k + b)^2 - b^2}  -  \int \frac{\mathrm{d}^2k}{(2\pi)^2} \frac{e^{-k^2/2\sigma^2} (\eta_\mathrm{mono} k^2 + \eta_f k + b)}{(\eta_\mathrm{mono} k^2 + \eta_f k + b)^2 - b^2} \nonumber \\
&+& \frac{1}{L^2}  \sum_{\vec{k}}  \frac{(1-e^{-k^2/2\sigma^2}) (\eta_\mathrm{mono} k^2 + \eta_f k + b)}{(\eta_\mathrm{mono} k^2 + \eta_f k + b)^2 - b^2}  -  \int \frac{\mathrm{d}^2k}{(2\pi)^2} \frac{(1-e^{-k^2/2\sigma^2}) (\eta_\mathrm{mono} k^2 + \eta_f k + b)}{(\eta_\mathrm{mono} k^2 + \eta_f k + b)^2 - b^2} \Bigg]
\label{eq:dT_monotopic_solution}
\end{eqnarray}
where $\eta_{\mathrm{mono}}=\eta_m/2$ is the monolayer surface viscosity and $A(k)$ and $B(k)$ as defined in the main text. For finite $\sigma$, the last line is close to zero because it is a Riemann sum minus the corresponding integral. For $\sigma\rightarrow\infty$, the last line vanishes exactly. In this way, we remove the singularity. The first line and the first term of the second line can be evaluated numerically the same way as for the transmembrane case.
We still have to evaluate the remaining integral
\begin{equation}
I = \int \frac{\mathrm{d}^2k}{(2\pi)^2} \frac{e^{-k^2/2\sigma^2} (\eta_\mathrm{mono} k^2 + \eta_f k + b)}{(\eta_\mathrm{mono} k^2 + \eta_f k + b)^2 - b^2}
\end{equation}
which we rewrite in polar coordinates 
\begin{equation}
I = \int \frac{\mathrm{d}k}{2\pi} \frac{k\,e^{-k^2/2\sigma^2} (\eta_\mathrm{mono} k^2 + \eta_f k + b)}{(\eta_\mathrm{mono} k^2 + \eta_f k + b)^2 - b^2}
\end{equation}
and for lack of an analytical solution evaluate numerically along with the other terms.
To extrapolate from numerical results for finite $\sigma$ to infinity, we take advantage of the approximately linear dependence on $1/\sigma^2$. Because $\tanh(x)\approx 1$ and $\sinh(x) \to \infty$ for large $x$, the summands decay exponentially for large $k$. Approximately, the limit is approached as $\Delta T^\mathrm{mono}(\sigma) \approx \Delta T + a/\sigma^2$. From two values at $\sigma_1$ and $\sigma_2$, we extrapolate 
\begin{equation}
\Delta T^\mathrm{mono} \approx \frac{\sigma_1^2 \Delta T^\mathrm{mono}(\sigma_1) - \sigma_2^2 \Delta T^\mathrm{mono}(\sigma_2)}{\sigma_1^2-\sigma_2^2} 
\end{equation}
and find numerical convergence for $\sigma_1 = 5\times 2\pi/L$ and $\sigma_2 = 6\times 2\pi/L$.

\subsection{Immersed-Boundary Method}

To predict the diffusion coefficient in an infinite and in a periodic system, Camley et al.~\cite{camley2015a} used the immersed boundary (IB) method. In this approach, the solid diffusing object is approximated as a fluid region to avoid the solution of boundary value problems. The Oseen tensor is modified by a Gaussian function whose width is chosen to reproduce results of an extended SD model \cite{saffman1975a,hughes1981a,petrov2008a}. They arrive at the following expressions for transmembrane inclusions (with $\beta=0.828494$)~\cite{camley2015a}:
\begin{eqnarray}
\label{eq:ib-trans-pbc}
D_\mathrm{PBC}^\mathrm{trans} &=& \frac{k_\mathrm{B}T}{2L^2} \sum_{\vec{k}(\neq 0)} \frac{1}{\eta_m k^2 + 2\eta_f k \tanh(kH)} \exp\left(-\frac{1}{2}k^2 \beta^2 R^2\right) \\[1ex]
\label{eq:ib-trans-inf}
D_0^\mathrm{trans} &=& \frac{k_\mathrm{B}T}{2} \int \frac{\mathrm{d}^2 k}{(2\pi)^2} \frac{1}{\eta_m k^2 + 2\eta_f k} \exp\left(- \frac{1}{2} k^2 \beta^2 R^2 \right) 
\end{eqnarray}
For monotopic inclusions, the IB method gives~\cite{camley2015a}:
\begin{eqnarray}
\label{eq:ib-mono-pbc}
D_\mathrm{PBC}^\mathrm{mono} &=& \frac{k_\mathrm{B}T}{2L^2} \sum_{\vec{k}(\neq 0)} \frac{A(k)}{A(k)^2-B(k)^2} \exp\left(-\frac{1}{2}k^2 \beta^2 R^2\right) \\[1ex]
\label{eq:ib-mono-inf}
D_0^\mathrm{mono} &=& \frac{k_\mathrm{B}T}{2} \int \frac{\mathrm{d}^2 k}{(2\pi)^2} \left[\frac{A(k)}{A(k)^2-B(k)^2} \right]_{H\rightarrow\infty} \exp\left(- \frac{1}{2} k^2 \beta^2 R^2 \right)\\[2ex]
A(k) &\rightarrow & \eta_\mathrm{mono} k^2 + \eta_f k + b \qquad\mathrm{for}\quad H\rightarrow\infty\\[2ex]
B(k) &\rightarrow & b\qquad\mathrm{for}\quad H\rightarrow\infty 
\end{eqnarray}
with $A(k)$ and $B(k)$ as defined in the main text.

The IB approach  \cite{camley2015a} is able to calculate a prediction for both the infinite-system value as well as the value under PBC. It also takes the explicit dependence on the radius into account. However, we should keep in mind that the Oseen tensor itself is only an approximation for the case in which the distance between two particles is much larger than the particle size.

\subsection{Comparison of Oseen and IB Theoretical Descriptions}

\begin{figure}[ht]
	 \centering
	 \includegraphics[width=1.0\linewidth]{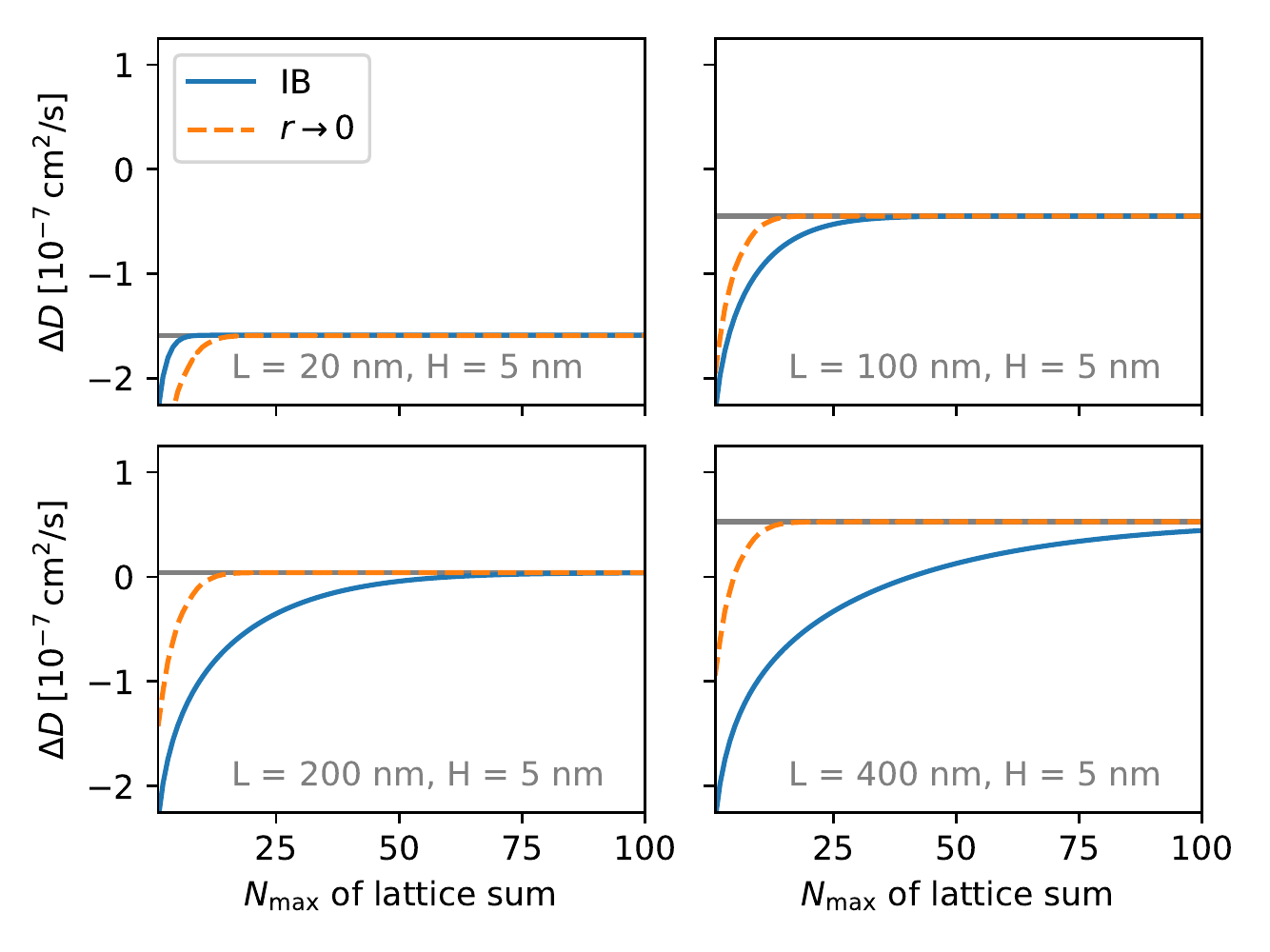} 
         \caption{Theoretical prediction of the difference $\Delta D=D_{\mathrm{PBC}}-D_0$ between the diffusion coefficients in PBC and in the infinite system depending on the number of summands (${n_x}^2+{n_y}^2\le {N_{\mathrm{max}}}^2$) considered in the lattice sum for four different box geometries (constant $H=5\,\mathrm{nm}$ and $L=20$, 100, 200, and 400 nm) and a protein radius of $1\,\mathrm{nm}$.
           The differences were calculated for the Oseen correction (orange lines) evaluated using Eq.~(\ref{eq:dT_transmembrane_solution}) ($r\rightarrow 0$), and for the IB method (blue lines) using the difference of Eq.~(\ref{eq:ib-trans-pbc}) and~Eq.~(\ref{eq:ib-trans-inf}) (IB). The gray horizontal line shows the converged value.}
     \label{fig:convergence_comparison}
\end{figure}

When calculating the lattice sums for the theoretical descriptions, care should be taken that convergence is reached. For the IB method, convergence depends on the box geometry while we can safely assume the sums in the point-perturbation method to be converged after 20-30 summands (Fig.~\ref{fig:convergence_comparison}). 

\begin{figure}[ht]
	 \centering
	 \includegraphics[width=0.48\linewidth]{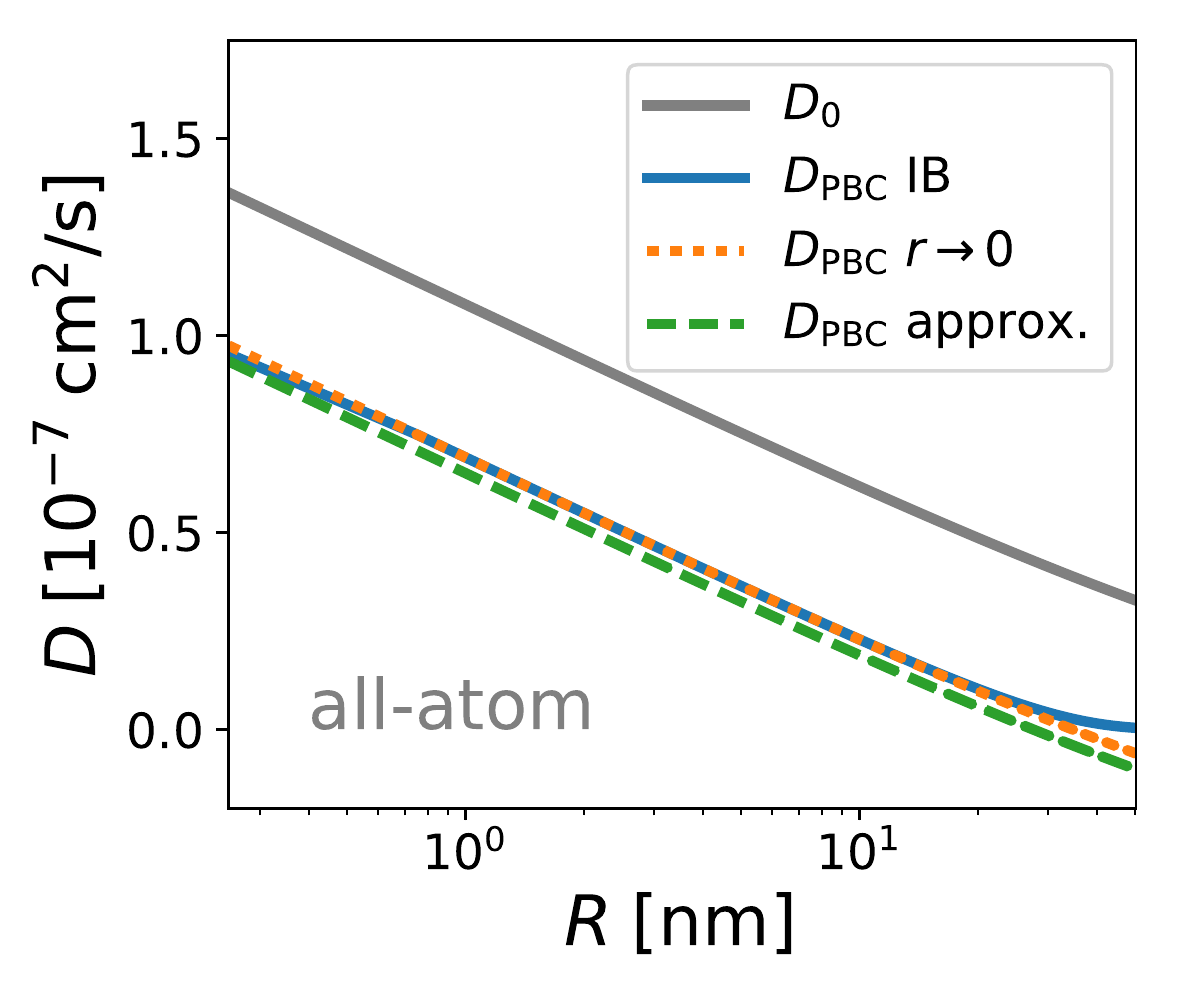} 
	 \includegraphics[width=0.48\linewidth]{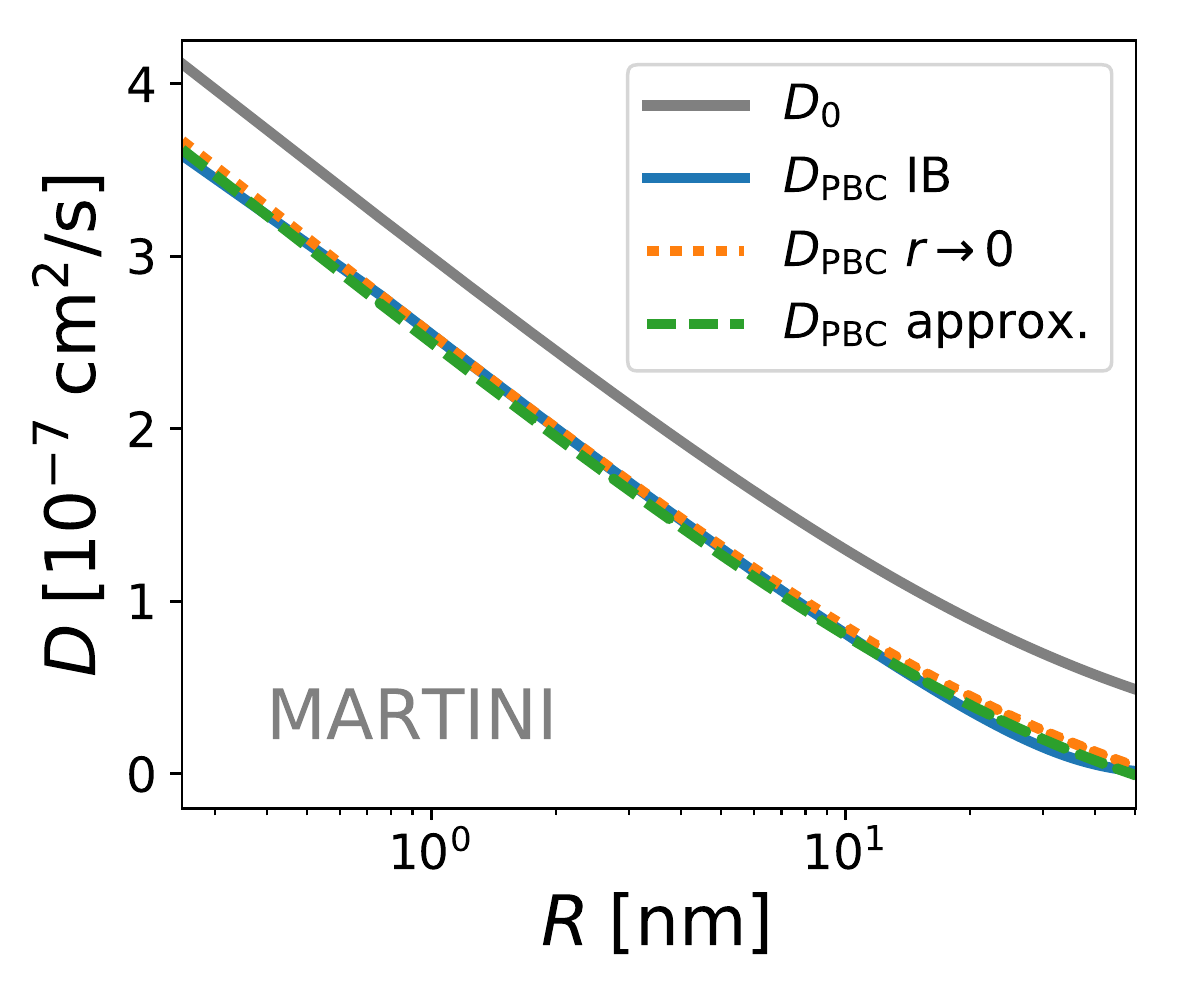} 
      \caption{Theoretical prediction of the diffusion coefficient of a cylindrical membrane inclusion as a function of its radius $R$ for different levels of theory for properties of typical all-atom (left) and MARTINI (right) simulations. Values for all-atom: $\eta_f=5.47\times 10^{-4}\;\mathrm{Pa\,s}$, $\eta_m=1.6\times 10^{-10}\;\mathrm{Pa\,s\,m}$. Values for MARTINI: $\eta_f=7.0\times 10^{-4}\;\mathrm{Pa\,s}$, $\eta_m=4.0\times 10^{-11}\;\mathrm{Pa\,s\,m}$. In both cases, $L=100\,\mathrm{nm}$ and $H=5\,\mathrm{nm}$. Results are shown for the Oseen correction (orange lines), the approximate Oseen correction Eq.~(2) (green lines), and the IB method (blue lines). For reference, the gray lines show $D_0$ calculated from Eq.~(\ref{eq:ib-trans-inf}). Both in the all-atom and MARTINI systems, the Oseen and IB corrections coincide almost perfectly.  Diffusion coefficients in PBC were calculated according to Eq.~(\ref{eq:ib-trans-pbc}) (IB) and by adding to Eq.~(\ref{eq:ib-trans-inf}) the corrections from Eq.~(\ref{eq:dT_transmembrane_solution}) ($r\rightarrow 0$) and Eq.~(2) (flat-box approximation), respectively. The deviations shown for very small radii are caused by the slower convergence of the IB method (see Fig.~\ref{fig:convergence_comparison}).}
     \label{fig:hydrodynamic_radius_comparison}
\end{figure}

We calculated the radius dependence of those two theories and for the flat-box approximation (Eq.~(2)) for typical values of fully atomistic and MARTINI coarse-grained simulations (Fig.~\ref{fig:hydrodynamic_radius_comparison}). 

For the monotopic correction, we compare the descriptions around two typical regions of the interleaflet friction coefficient $b$ (Fig.~\ref{fig:interleaflet_coupling_comparison}).
Typical values are for all-atom simulations ($b\approx10^8\,\mathrm{Pa\,s/m}$) and for MARTINI ($b\approx10^6-10^7\,\mathrm{Pa\,s/m}$) \cite{camley2015a}.

\subsection{Comparison of Monotopic and Transmembrane Correction}

\begin{figure}[ht]
	 \centering
	 \includegraphics[width=0.49\linewidth]{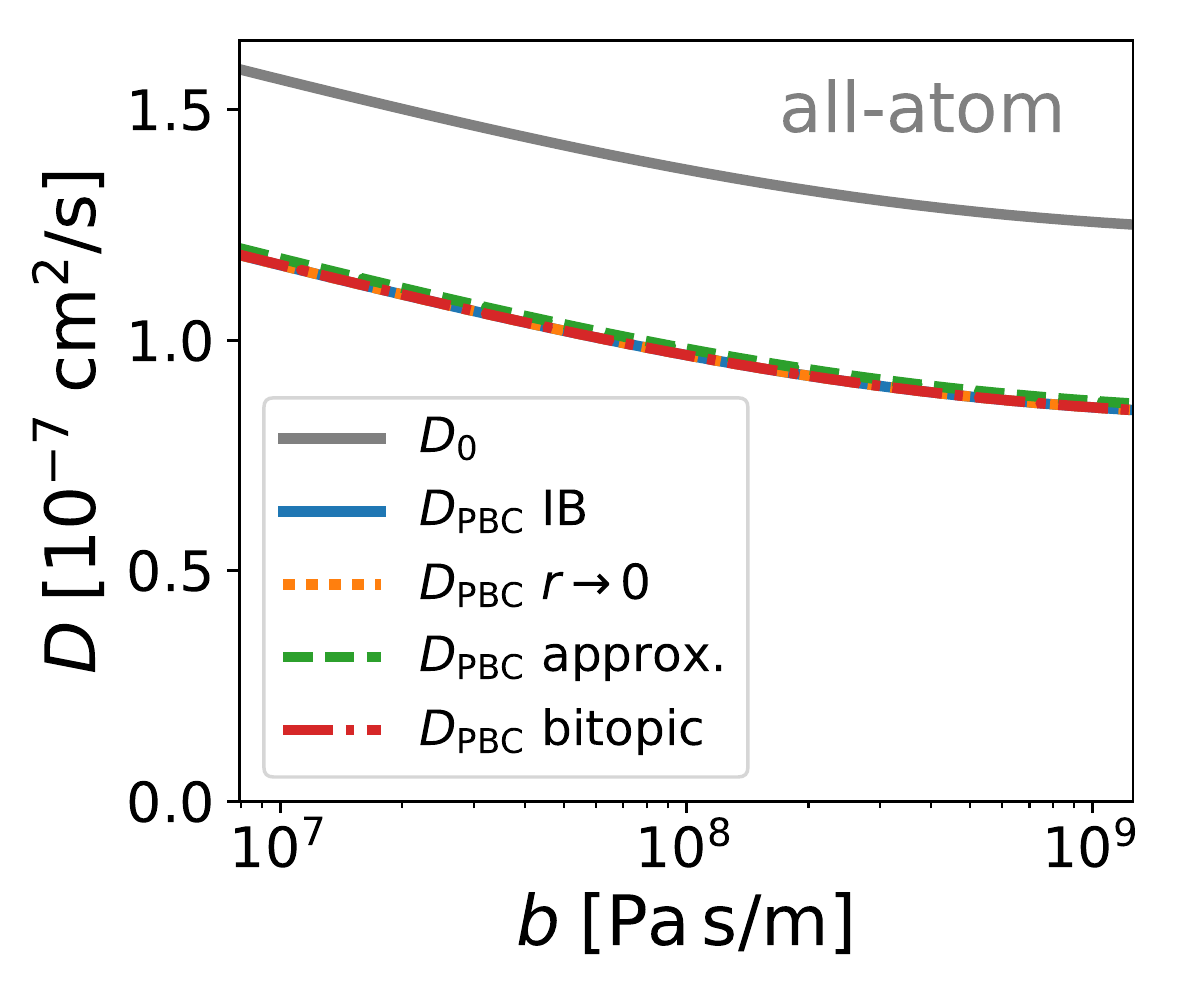} 
	 \includegraphics[width=0.49\linewidth]{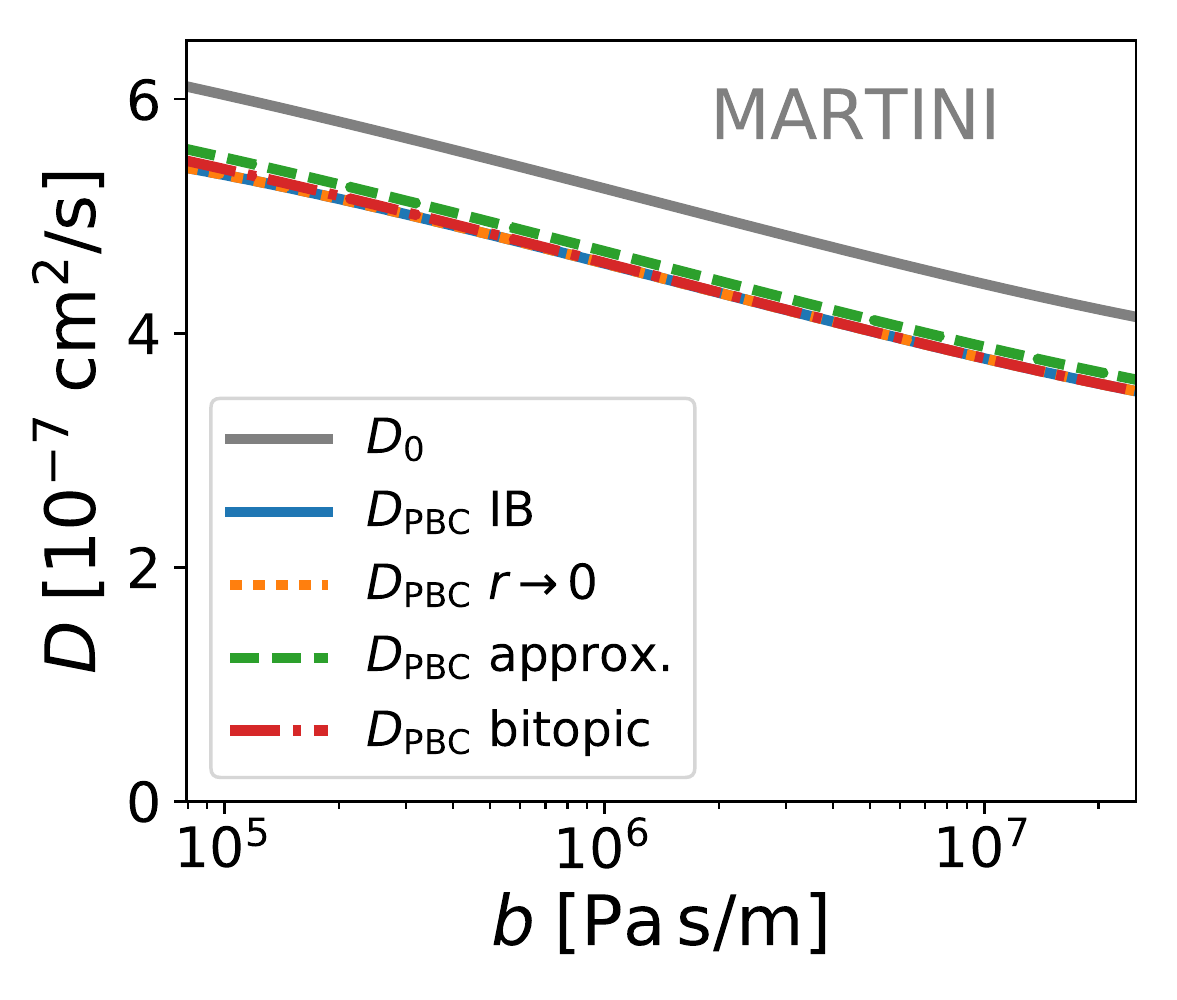} 
         \caption{Theoretical prediction of the diffusion coefficient of a cylindrical monotopic membrane inclusion with radius 0.5~nm as a function of interleaflet friction coefficient $b$ for properties of typical all-atom (left) and MARTINI (right) simulations. Values for all-atom: $\eta_f=5.47\times 10^{-4}\;\mathrm{Pa\,s}$, $\eta_m=1.6\times 10^{-10}\;\mathrm{Pa\,s\,m}$. Values for MARTINI: $\eta_f=7.0\times 10^{-4}\;\mathrm{Pa\,s}$, $\eta_m=4.0\times 10^{-11}\;\mathrm{Pa\,s\,m}$. In both cases, $L=100\,\mathrm{nm}$ and $H=5\,\mathrm{nm}$. Results are shown for the Oseen correction (orange lines), the approximate Oseen correction Eq.~(2) (green lines), and the IB method (blue lines). For reference, the gray lines show $D_0$ calculated from Eq.~(\ref{eq:ib-mono-inf}). Both in the all-atom and MARTINI systems, the Oseen and IB corrections coincide almost perfectly. Diffusion coefficients in PBC were calculated according to Eq.~(\ref{eq:ib-mono-pbc}) (IB) and by adding to Eq.~(\ref{eq:ib-mono-inf}) the corrections from Eq.~(\ref{eq:dT_monotopic_solution}) ($r\rightarrow 0$) and Eq.~(2) (flat-box approximation), respectively.}
     \label{fig:interleaflet_coupling_comparison}
\end{figure}

In comparisons for typical membrane physical parameters (such as in Fig.~\ref{fig:interleaflet_coupling_comparison}), we found the difference in the monotopic and transmembrane corrections to be negligible, especially for the high intermonolayer friction in atomistic simulations. Notable deviations are only observed for $b < 10^6$ Pa~s/m, a regime just below the one of MARTINI simulations.
The weak dependence on the interleaflet friction coefficient $b$ renders it impossible to obtain $b$ from the size dependence. As an advantage, the use of the transmembrane (bitopic) correction and its flat-box approximation also for monotopic molecules simplifies the analysis.

For large heights, the monotopic and bitopic (transmembrane) expressions converge. For $H \rightarrow\infty$ and $k$ small but finite, we have for the reciprocal of the summand in the monotopic case:
$(A^2(k)-B^2(k))/A(k) \approx 2 \eta_f k + ( \eta_m - \eta_f^2/b ) k^2 + \mathcal{O}(k^3)$.
For the bitopic case, we have exactly $2 \eta_f k + \eta_m k^2$. 
That is, the two expressions have the same leading term and almost the same secondary term. The relative deviation of the factor of $k^2$ is therefore $\eta_f^2/(\eta_m b) \approx 1/40$. 
With MARTINI values, this means that also the $k^2$ term is essentially identical with only a 2.5~\% correction. We therefore expect the bitopic correction to be sufficient in most cases and interpret the term $\eta_f^2/(\eta_m b)$ as the (dimensionless) relative importance of using the monotopic correction instead of the bitopic one.
For our POPC simulations, we estimate an importance of $\eta_f^2/(\eta_m b) \approx 0.008$ for the monotopic correction from the results of the fit and from a value of $b = 2.8\times 10^6$ Pa~s/m. This value was estimated for a similar lipid in the MARTINI model, however with a longer saturated tail (five beads instead of four) and at 323~K \cite{denotter2007a}, but we assume it provides a good estimate and we can use it to compare the two formalisms

\begin{figure}[ht]
	 \centering
	 \includegraphics[width=0.49\linewidth]{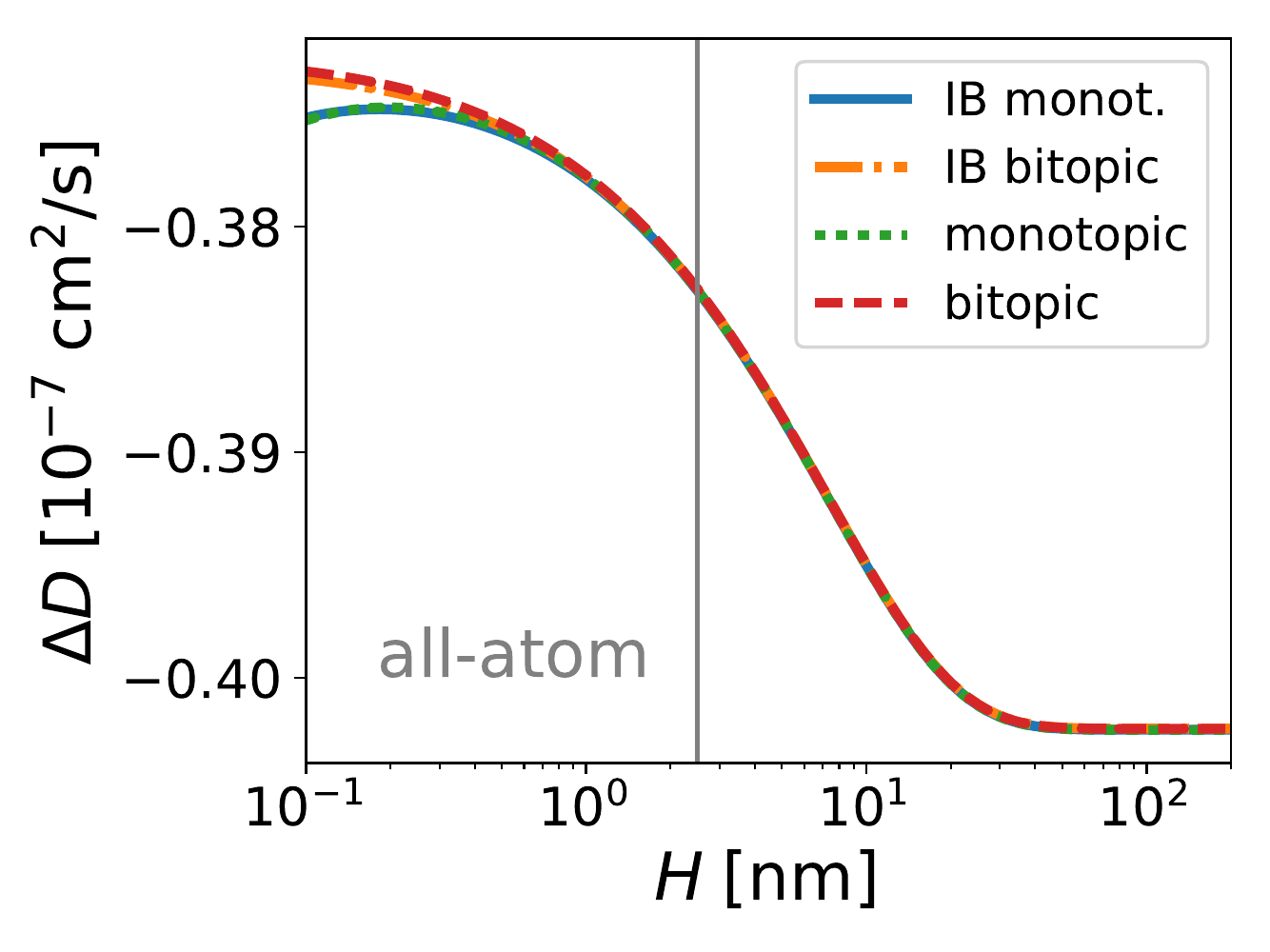} 
	 \includegraphics[width=0.49\linewidth]{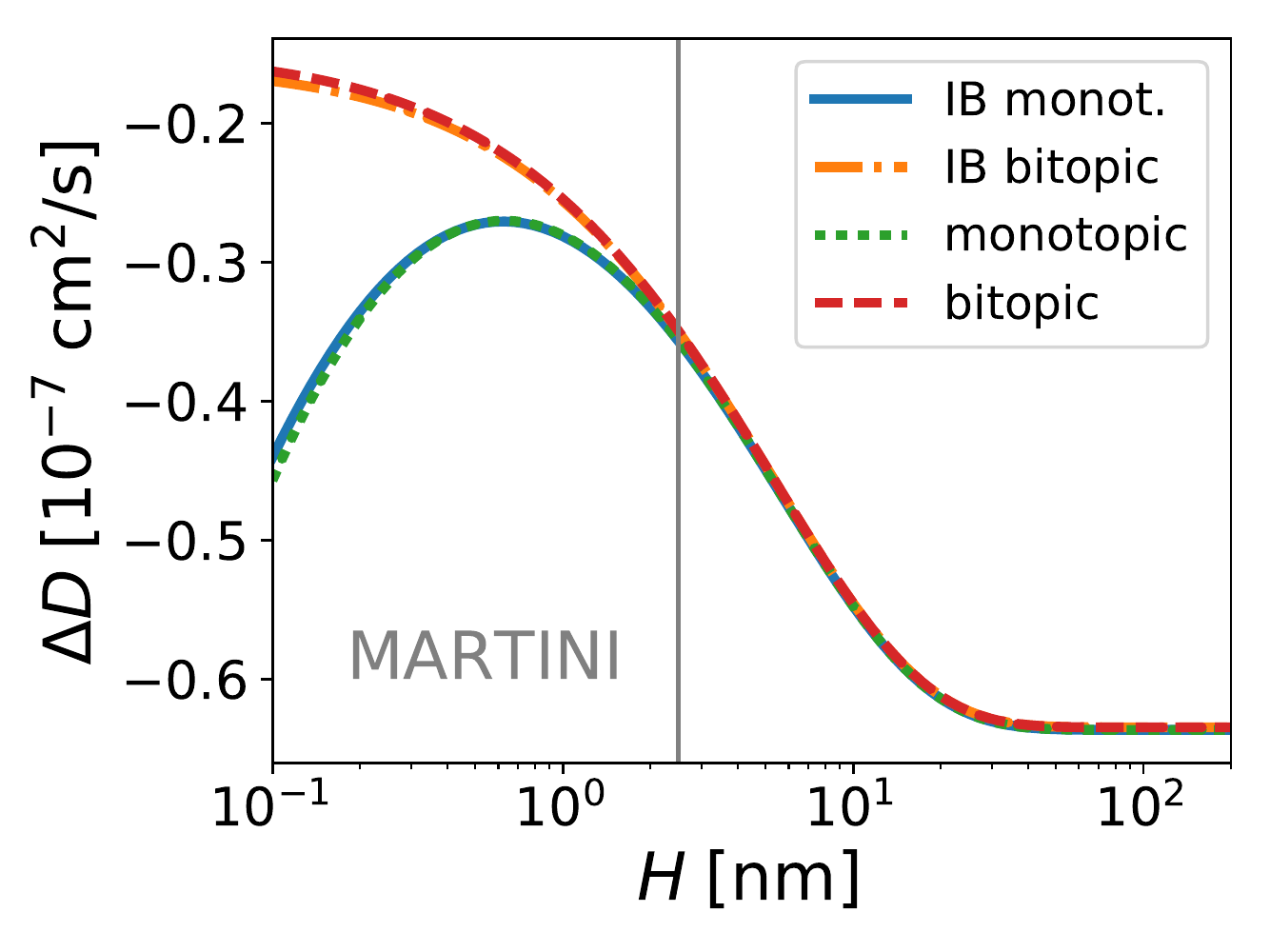} 
         \caption{Theoretical prediction of the difference $\Delta D=D_{\mathrm{PBC}}-D_0$ between the diffusion coefficients in PBC and in the infinite system for a cylindrical monotopic membrane inclusion with radius 0.5~nm as a function of $H$ for properties of typical all-atom (left) and MARTINI (right) simulations. Values for all-atom: $\eta_f=5.47\times 10^{-4}\;\mathrm{Pa\,s}$, $\eta_m=1.6\times 10^{-10}\;\mathrm{Pa\,s\,m}$, $b=1.0\times10^8\;\mathrm{Pa\,s/m}$. Values for MARTINI: $\eta_f=7.0\times 10^{-4}\;\mathrm{Pa\,s}$, $\eta_m=4.0\times 10^{-11}\;\mathrm{Pa\,s\,m}$, $b=2.8\times10^6\;\mathrm{Pa\,s/m}$. In both cases, $L=100\,\mathrm{nm}$. Results are shown for the Oseen correction (green lines for monotopic and red lines for bitopic) and the IB method (blue lines for monotopic and orange lines for bitopic).  For large $H$, all corrections coincide almost perfectly. Vertical lines indicate a typical value in membrane simulations, $H=2.5$ nm.}
     \label{fig:monotopic_bitopic_h_dependence}
\end{figure}

In the monotopic IB and Oseen corrections for very small $H$, we noticed a turnover to a pathological divergence to $\Delta D \rightarrow -\infty$ (see Fig.~\ref{fig:monotopic_bitopic_h_dependence}). This small-$H$ divergence might be associated with the not strictly $z$-periodic formulation of the monotopic Oseen tensor \cite{camley2015a}. It is difficult to test from our data gained in a $b$-insensitive regime whether the monotopic Oseen tensor describes the dynamics significantly better than the standard bitopic one.

\clearpage
\newpage

\section{Simulation Details}

\subsection{Height Study}

\begin{figure}[ht]
	 \centering
	 \includegraphics[width=0.75\linewidth]{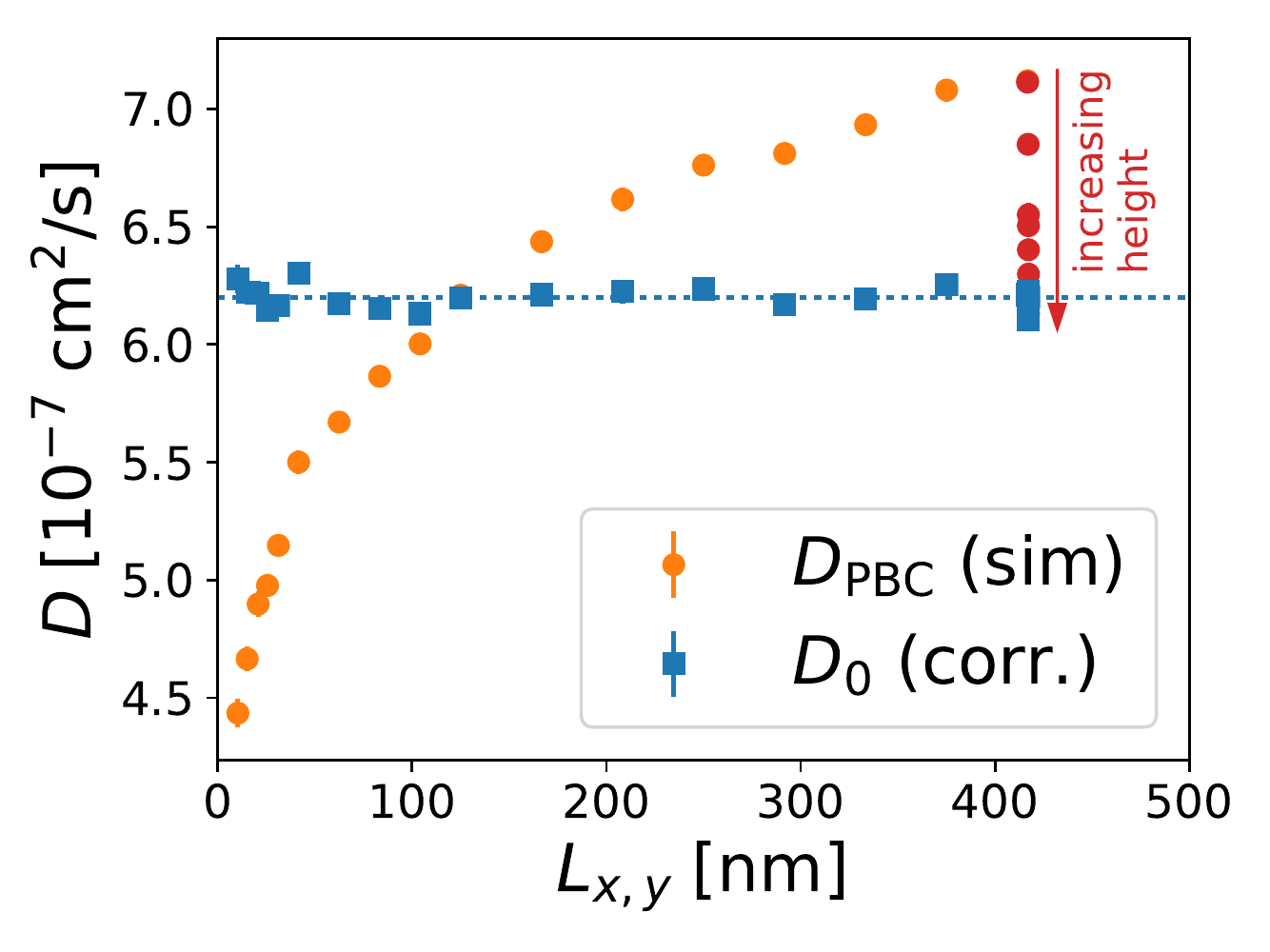}
         \caption{Diffusion coefficients in the full data set of POPC simulations with varied width (orange) at $H=2.25\,\mathrm{nm}$ and varied height (red) at $L=417\,\mathrm{nm}$, respectively, together with the corrected values and the best-fit infinite-system values (blue). The corresponding height dependence (here: red) for 417~nm is shown explicitly in Figure~1.
           \label{fig:full-height-highlighted}}
\end{figure}

\begin{figure}[ht]
	 \centering
	 \includegraphics[width=0.475\linewidth]{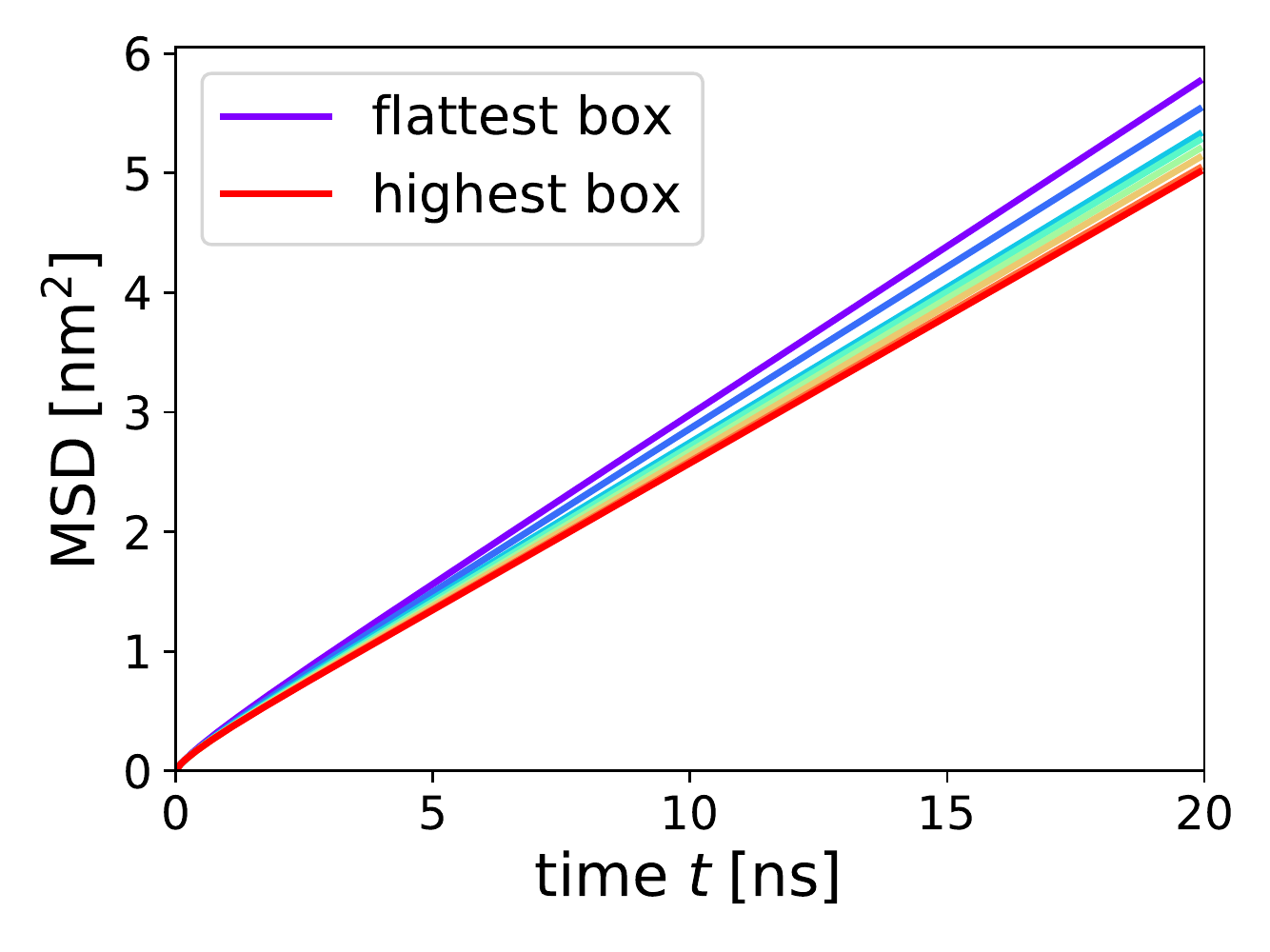}
	 \includegraphics[width=0.475\linewidth]{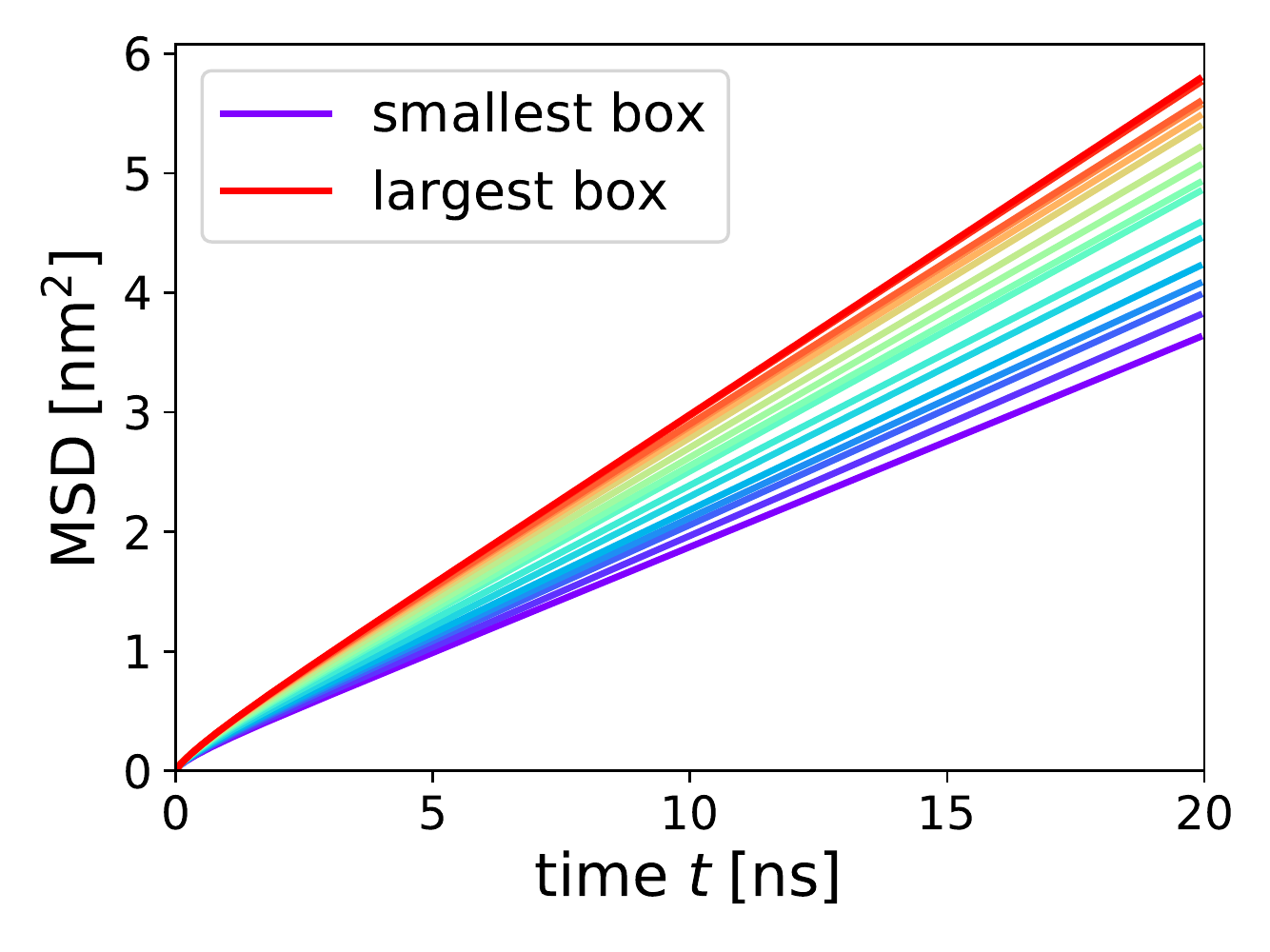}
	 \includegraphics[width=0.475\linewidth]{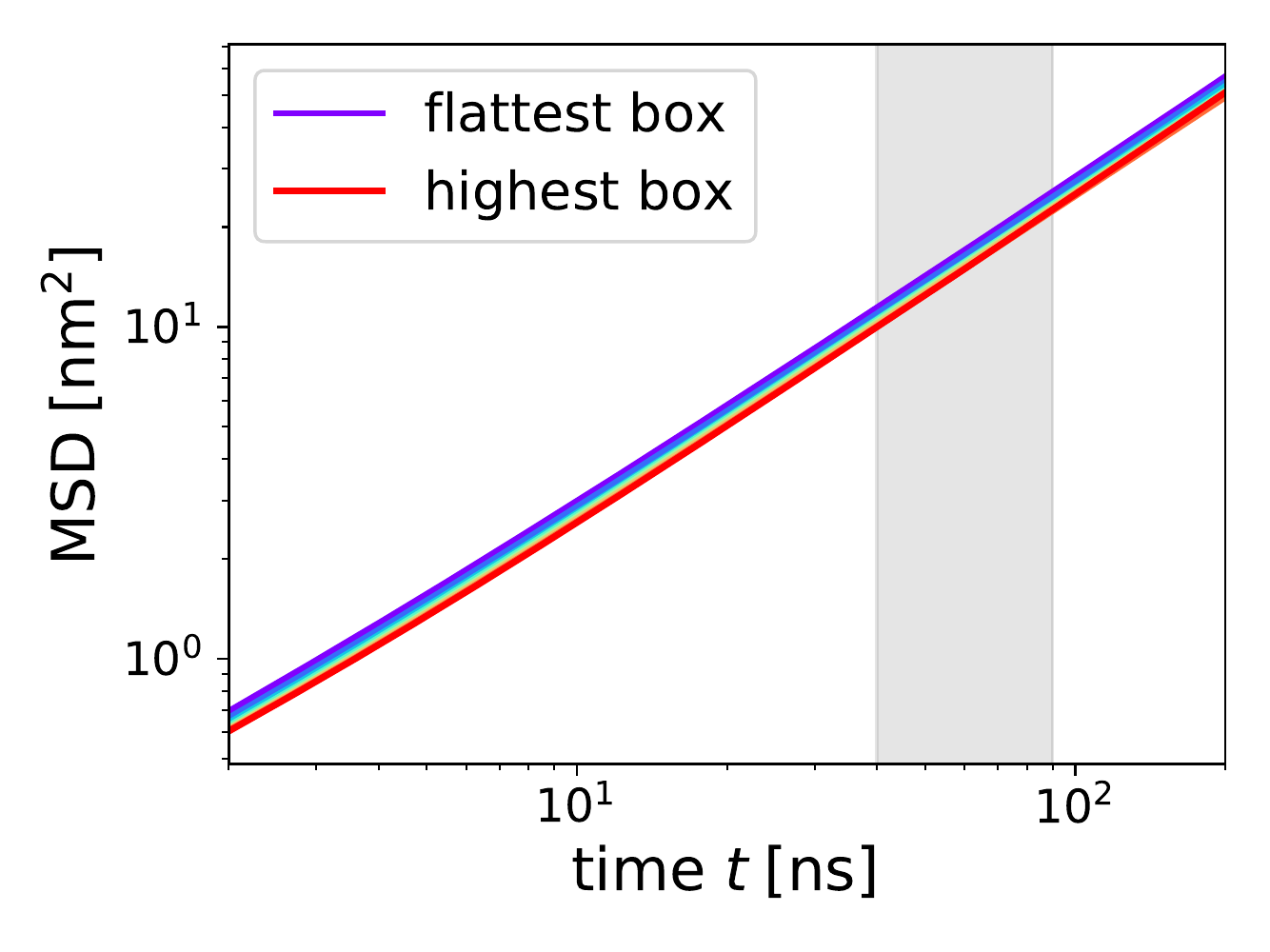}
	 \includegraphics[width=0.475\linewidth]{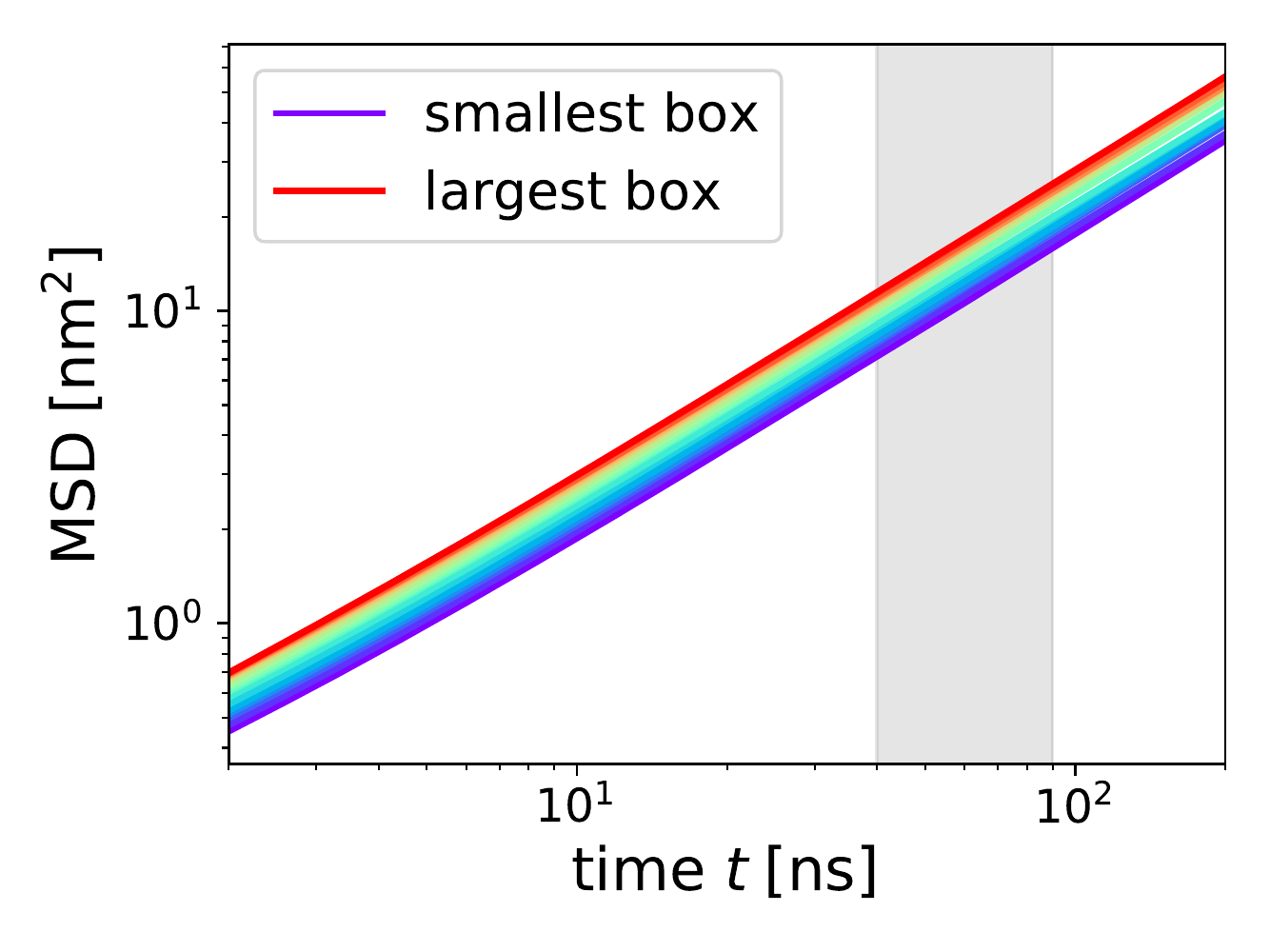}
         \caption{MSD curves for the height study (left) and, for comparison, from a width study (right) from earlier work \cite{voegele2016a}. The upper plots show the initial phase and the lower the long-time behavior on a logarithmic scale with the fitting range in gray.
           \label{fig:msd-study-H}}
\end{figure}

For our simulations of neat POPC membrane systems, we used the protocol and setup of \cite{voegele2016a}, except for variations in box geometry. To fit the viscosities and the infinite-system diffusion coefficient, we used the simulation data from both studies.
To give the context, we show the combined data in Fig.~\ref{fig:full-height-highlighted} and mean-squared displacement (MSD) curves of both studies in Fig.~\ref{fig:msd-study-H}.

\begin{table}[h]
\centering
\begin{tabular}{c|c|c|c|c|c}
init. $L_z$ [nm] & num. of lipids & num. of water beads & box width $\langle L \rangle$ [nm] & box height $\langle L_z \rangle$ [nm] & sim. time [$\mathrm{\mu s}$] \\ \hline
9  & 540800 &  6791600 & 416.71(5) & 9.43(1) & 2.00 \\
15 & 540800 & 14664800 & 417.02(1) & 15.44(1) & 2.00\\
22 & 540800 & 23850200 & 417.12(1) & 22.44(1) & 0.60\\
30 & 540800 & 34347800 & 417.17(1) & 30.46(1) & 2.00\\
40 & 540800 & 47469800 & 417.17(1) & 40.50(1) & 0.53\\
50 & 540800 & 60591800 & 417.20(1) & 50.49(1) & 2.00\\
75 & 540800 & 93396800 & 417.21(1) & 75.51(1) & 1.00\\
99 & 540800 &124889600 & 417.19(1) & 99.61(1) & 0.50
\end{tabular}
\caption{System parameters of the lipid membrane simulations to test the height dependence.} 
\label{tab:system-parameters-study-H}
\end{table}

In a typical application, one would use the viscosity $\eta_f$ of the bulk solvent as input instead of performing a computationally expensive height study. Using $\eta_f = 10.2$ Pa~s (as obtained from pressure fluctuations in bulk-solvent simulations; see Table \ref{tab:waterviscosity}) in the analysis of only the flat-box simulations of \cite{voegele2016a}, we obtained almost the same values of $D_0$ and $\eta_m$ for the POPC membrane as in the global fit that included the height-dependent simulations and left $\eta_f$ variable (main text). Results for the full Oseen correction Eq.~(1) using bitopic and monotopic Oseen tensors, and for the approximate bitopic correction Eq.~(2), are listed in Table~\ref{tab:correction-etaf-fixed}. Also included are fits with fixed $\eta_f$ to all POPC membrane simulations here and in \cite{voegele2016a}. The consistency of all fits justifies (1) fixing the solvent viscosity $\eta_f$ at a precalculated bulk value, and (2) using the approximate expression Eq.~(2) for the analysis of typical membrane simulations.

\begin{table}[h]
\centering
\begin{tabular}{c|c|c|c}
  data set    & correction  & $D_0$ [$10^{-7}$ cm$^2$/s] & $\eta_m$ [10$^{-11}$ Pa s m] ] \\ \hline
 flat boxes & monotopic   &  6.20(2) &  4.07(6) \\
 flat boxes & bitopic     &  6.16(2) &  3.94(6) \\
 flat boxes & approximate &  6.23(2) &  3.92(6) \\
 all data   & monotopic   &  6.20(2) &  4.06(6) \\
 all data   & bitopic     &  6.18(2) &  3.96(6) \\
 all data   & approximate &  6.20(2) &  3.89(6) \\
\end{tabular}
\caption{Infinite-system diffusion coefficient and membrane surface viscosity obtained from fits of the full Oseen correction Eq.~(1) using monotopic and bitopic Oseen tensors, and of the approximate bitopic Oseen correction Eq.~(2) at fixed fluid viscosity $\eta_f=10.2$ Pa~s. For reference, a bitopic fit to all data with $\eta_f$ variable gave (main text) $D_0 = 6.20(2)\times 10^{-7}$ cm$^2$/s and $\eta_m = 3.97(6)\times 10^{-11}$ Pa~s~m} 
\label{tab:correction-etaf-fixed}
\end{table}

\clearpage
\newpage

\subsection{Viscosity of Water from Pressure Fluctuations}

The viscosity of a specific water model under certain conditions can be obtained via the fluctuations of the off-diagonal elements of the pressure tensor in a simulation with constant box volume. 
We calculated $\eta_f$ according to the following Einstein relation~\cite{hess2002a}:
\begin{eqnarray}
\eta_f &=& \lim_{t\rightarrow\infty} \frac{1}{2} \frac{V}{k_{\mathrm{B}} T} \frac{\mathrm{d}}{\mathrm{d}t} \left\langle \left(  \int_{t_0}^{t_0+t} P_{xz}(t')\,\mathrm{d}t'  \right)^2 \right\rangle_{t_0} \label{eq:etafromp}
\end{eqnarray}
where $V$ is the box volume, $P_{xz}$ is any off-diagonal element of the pressure tensor, and $t$ is the time.

We simulated cubes of 524880 MARTINI water beads (10\% of which were antifreeze particles) for $1.0\;\mathrm{\mu s}$ at a fixed edge length of $40\,\mathrm{nm}$ with the same thermostat settings as in the respective membrane simulations. From these simulations, we obtained the values for the water viscosity by fitting the slope of the correlation function in Eq.~(\ref{eq:etafromp}) over the time range 64 to 80 ns.
For the temperatures and thermostats used in this work, we obtained the following MARTINI water viscosities: 
\begin{itemize}
\item $10.2(4) \;\mathrm{Pa\,s}$ for $300\;\mathrm{K}$ with the Berendsen thermostat \cite{berendsen1984a} according to the simulation protocol in \cite{voegele2016a}
\item $8.4(4) \;\mathrm{Pa\,s}$ for $310\,\mathrm{K}$ with the modified Berendsen thermostat for canonical sampling (Bussi-Donadio-Parrinello thermostat) \cite{bussi2007a}  according to the simulation protocol in \cite{hedger2016a} for the ANT1 protein-loaded system
\item $7.1(4) \;\mathrm{Pa\,s}$ for $323\;\mathrm{K}$ with the Bussi-Donadio-Parrinello thermostat \cite{bussi2007a} according to the simulation protocol in \cite{ingolfsson2014a} for the plasma membrane
\end{itemize}

\clearpage
\newpage

\subsection{Details on the Influence of the Water Viscosity}

In the simulations testing the influence of the water viscosity, we varied the mass of the MARTINI water beads and left all other parameters the same. The viscosity of bulk water scales with the square root of the water mass $M$, $\eta_f(M)=(M/M_0)^{1/2}\eta_f(M_0)$.  By contrast, we expect the surface viscosity $\eta_m$ to be relatively unaffected, such that $L_{\mathrm{SD}}\propto M^{-1/2}$ approximately.
We performed four simulations using MARTINI water beads with masses of $M=M_0=72$ and $M=144$, 288 and 720~amu. The box height was $L_z=9$~nm, and the width $L=40$~nm.


\begin{figure}[ht]
 	 \includegraphics[width=0.45\linewidth]{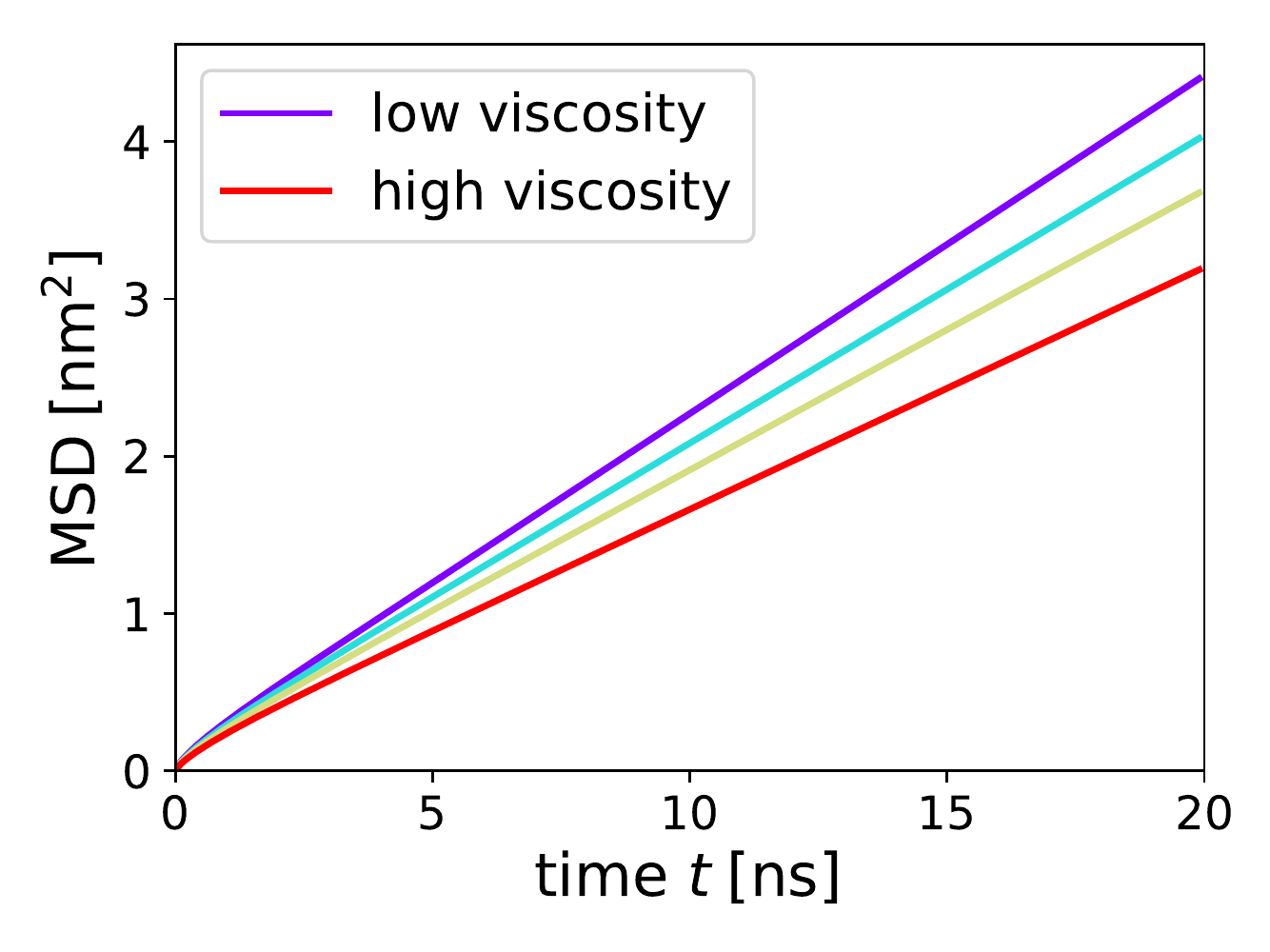} 
 	 \includegraphics[width=0.45\linewidth]{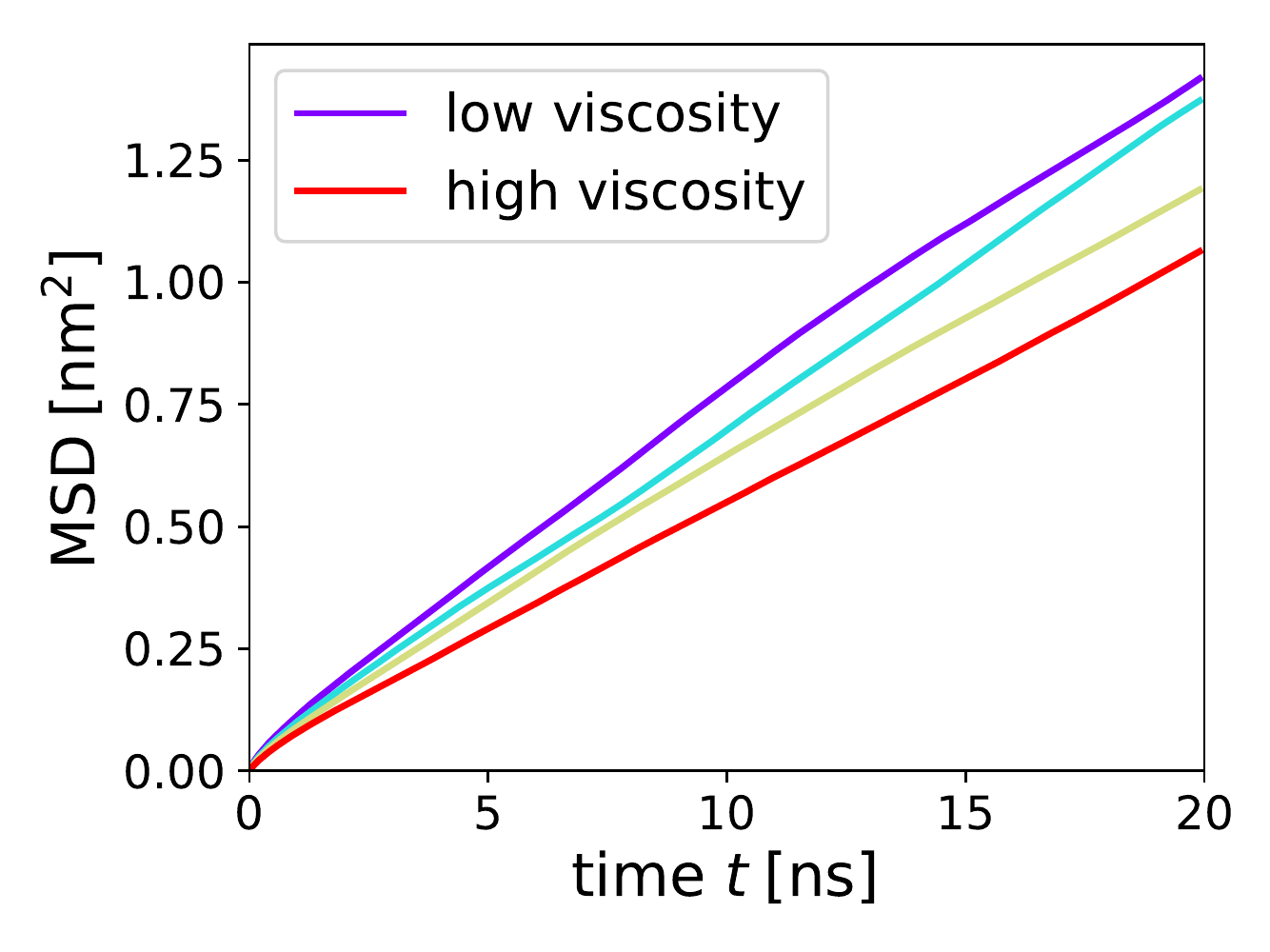}
 	 \includegraphics[width=0.45\linewidth]{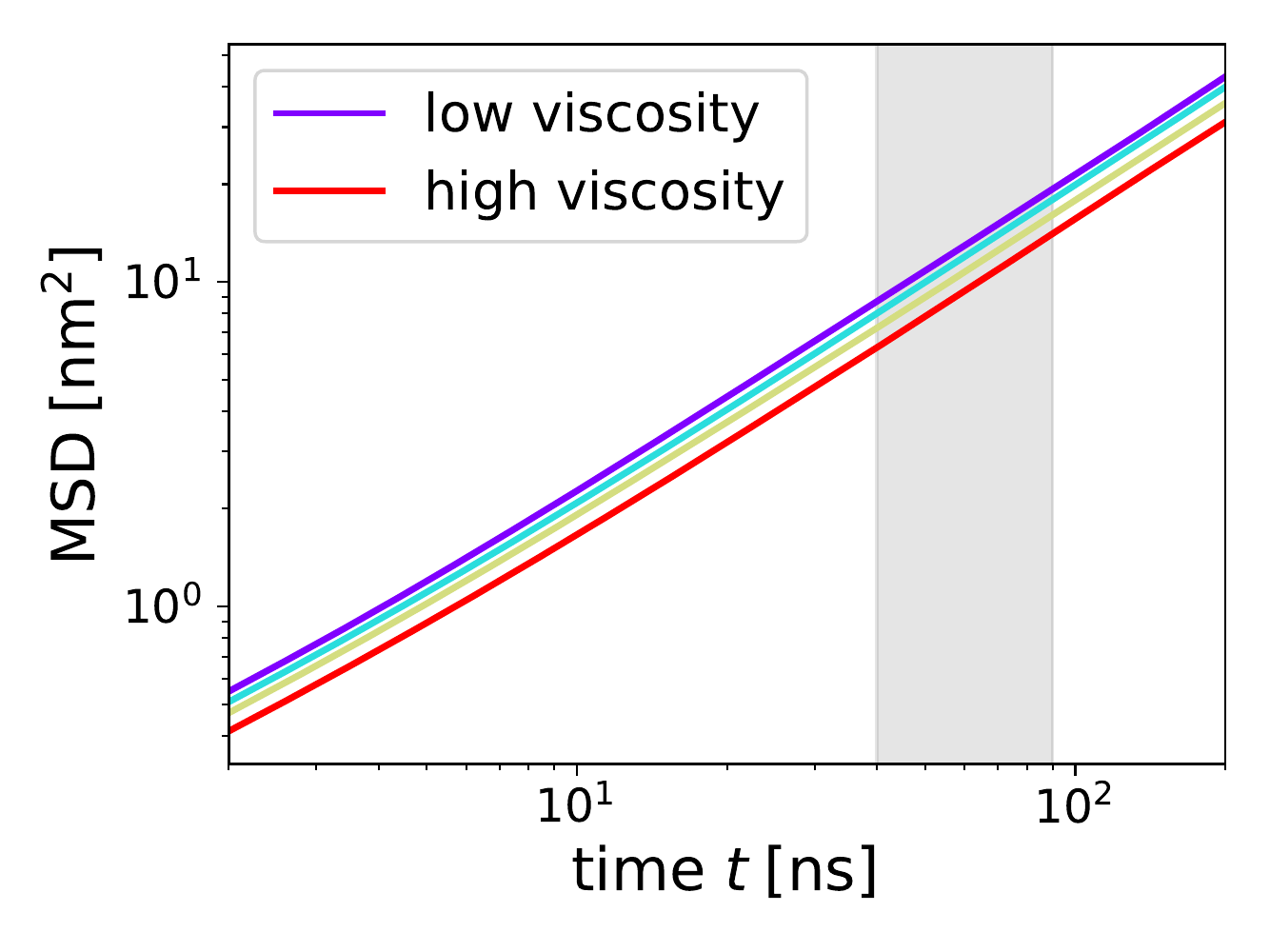} 
 	 \includegraphics[width=0.45\linewidth]{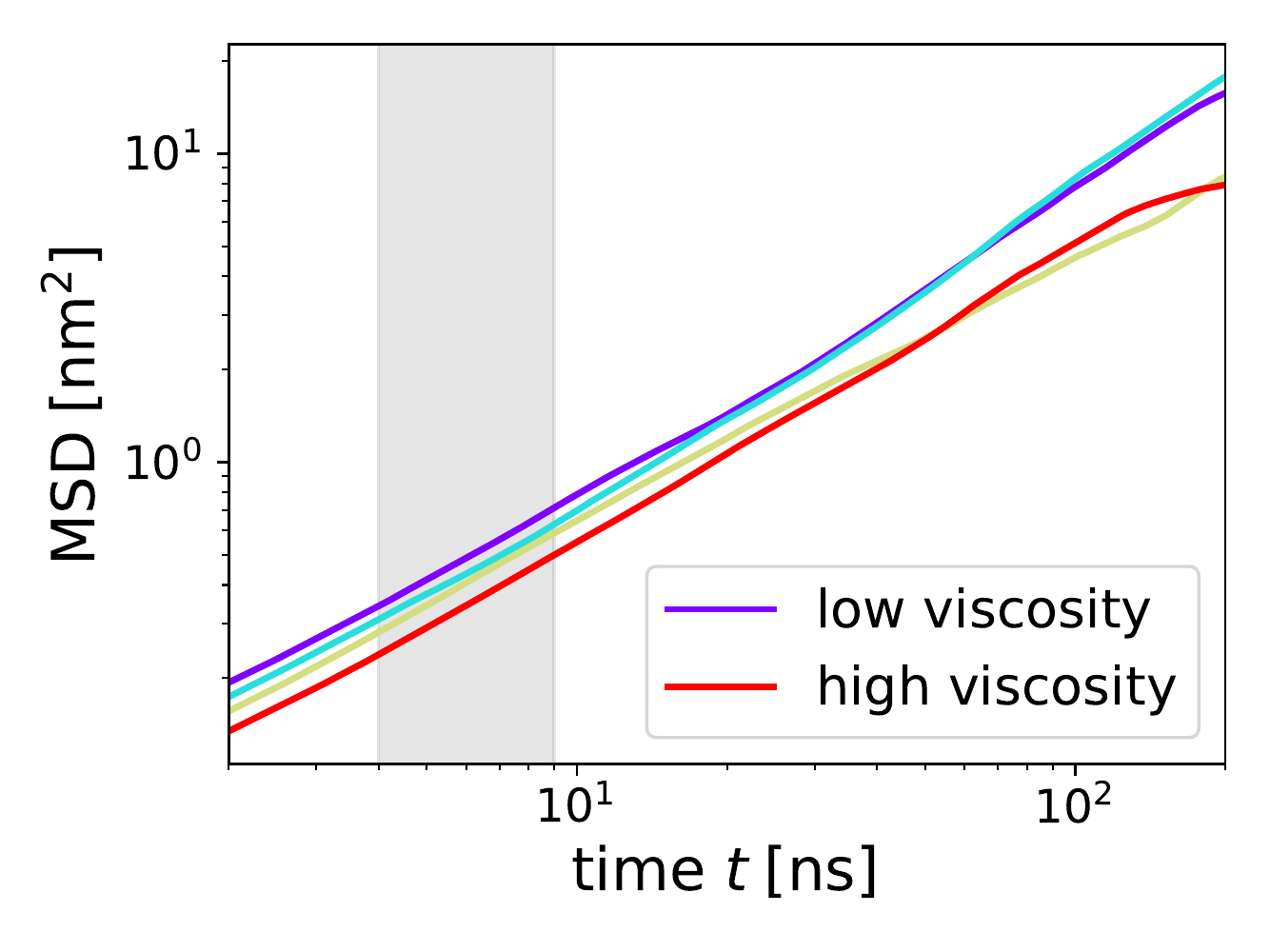}
          \caption{ \label{fig:msd-waterviscosity} MSD of POPC lipids (left) and membrane-spanning CNTs \cite{voegele2016a} (right) for different water viscosities in a box of size $L=40$~nm and $L_z=9$~nm. The viscosity $\eta_f$ of MARTINI water was varied by changing the mass of water particles from 72 amu to 144, 288, and 720 amu, respectively. The upper plots show the initial phase and the lower the long-time behavior on a logarithmic scale with the fitting range in gray. }
\end{figure}

\begin{table}[ht]
\begin{center}
\begin{tabular}{c|c|c|c|c}
$M\;\mathrm{[amu]}$ & 72 &  144 &  288 & 720 \\ \hline
$\eta_f\;[\mathrm{10^{-4}\,Pa\,s}]$ & $10.2(2)$ & $14.4(2)$ & $20.4(2)$ & $32.2(2)$ \\ \hline 
$\Delta D/D_0$(POPC) & -12.5\,\% & -10.5\,\% &  -8.1\,\% &  -5.1\,\% \\ \hline
$\Delta D/D_0$(CNT)  & -29.2\,\% & -25.1\,\% & -19.6\,\% & -11.9\,\% 
\end{tabular}
\end{center}
\caption{Details on the various water viscosities: Mass $M$ of the water beads, corresponding water viscosity $\eta_f$, and resulting corrections on the diffusion coefficient for lipids and the CNT. }
\label{tab:waterviscosity}
\end{table}

\begin{table}[h]
\centering
\begin{tabular}{c|c|c|c|c|c}
W mass [amu] & num. of lipids & num. of water beads & box width $\langle L \rangle$ [nm] & box height $\langle L_z \rangle$ [nm] & sim. time [$\mathrm{\mu s}$] \\ \hline
72 & 5408 & 67916 & 41.69(5) & 9.43(2) & 2.00 \\
144 & 5408 & 67916 & 41.69(5) & 9.42(2) & 2.00 \\
288 & 5408 & 67916 & 41.69(5) & 9.43(2) & 2.00 \\
720 & 5408 & 67916 & 41.70(5) & 9.42(2) & 2.00 
\end{tabular}
\caption{System parameters of the lipid membrane simulations to test the viscosity dependence.} 
\label{tab:system-parameters-viscosity}
\end{table}

\clearpage
\newpage

\subsection{Details of the Protein-Crowded Membrane}

\begin{table}[h]
\centering
\begin{tabular}{c|c|c|c|c|c|c|c}
ANT1 & POPC & POPE & Card. & solvent & box width $\langle L \rangle$ [nm] & box height $\langle L_z \rangle$ [nm] & sim. time [$\mathrm{\mu s}$] \\ \hline
1 & 220 & 160 & 20 & 6588 & 12.05(5) & 10.19(9) & 2.00 \\
4 & 880 & 640 & 80 & 26332 & 24.10(5) & 10.19(4) & 2.00 \\
9 & 1980 & 1440 & 180 & 59202 & 36.15(5) & 10.18(3) & 2.00 \\
16 & 3520 & 2560 & 320 & 105440 & 48.20(5) & 10.19(2) & 2.00 \\
25 & 5500 & 4000 & 500 & 164875 & 60.24(5) & 10.20(2) & 2.00 \\
49 & 10780 & 7840 & 980 & 322812 & 84.34(5) & 10.19(1) & 2.00 \\
100 & 22000 & 16000 & 2000 & 658400 & 120.48(6) & 10.19(1) & 2.00 \\
225 & 49500 & 36000 & 4500 & 1480950 & 180.72(6) & 10.19(1) & 2.00 \\
400 & 88000 & 64000 & 8000 & 2634800 & 240.96(6) & 10.19(1) & 2.00 \\
625 & 137500 & 100000 & 12500 & 4111250 & 301.20(7) & 10.18(1) & 2.00 \\
900 & 198000 & 144000 & 18000 & 5925600 & 361.45(8) & 10.19(1) & 2.00
\end{tabular}
\caption{System parameters of the ANT1 simulations.} 
\label{tab:system-parameters-ant1}
\end{table}

\begin{figure}[ht]
	\centering
	$\qquad\qquad$ Protein \hspace{6cm} Cardiolipin (lower leaflet)\\
	\includegraphics[width=0.49\linewidth]{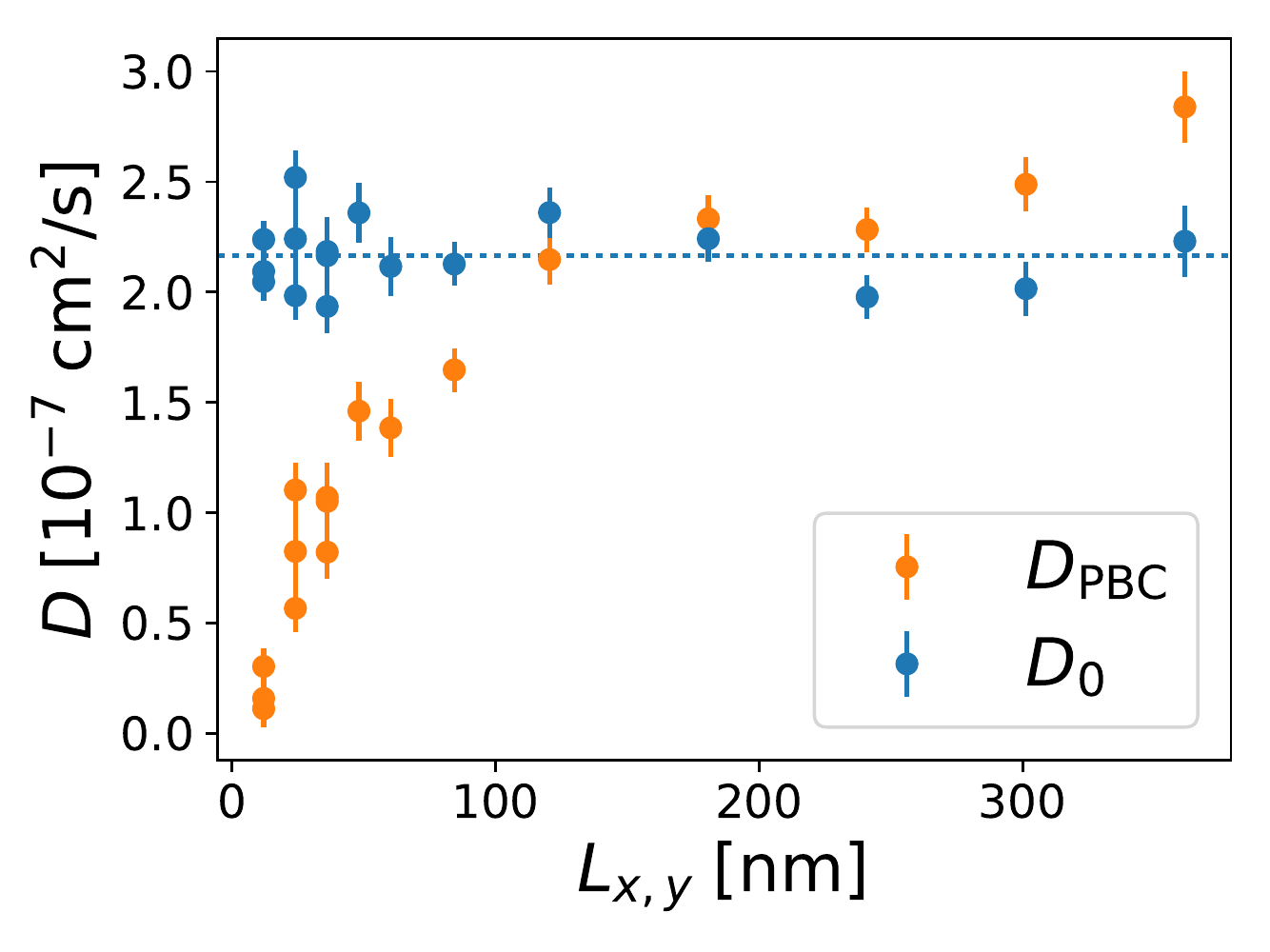}
	\includegraphics[width=0.49\linewidth]{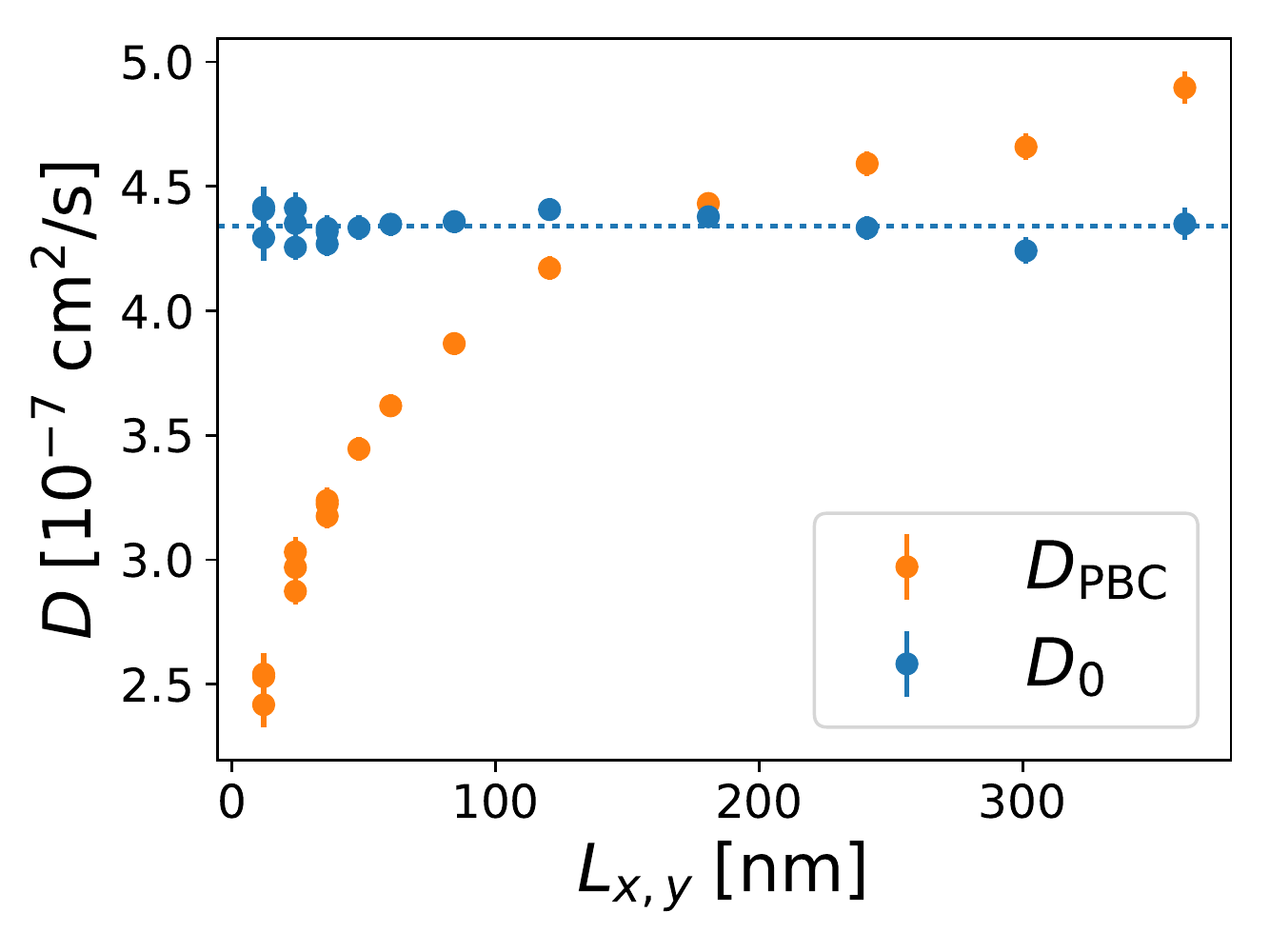}
	$\qquad$ POPC (upper leaflet) $\qquad\qquad\qquad\qquad\qquad\qquad\;$ POPC (lower leaflet)\\
	\includegraphics[width=0.49\linewidth]{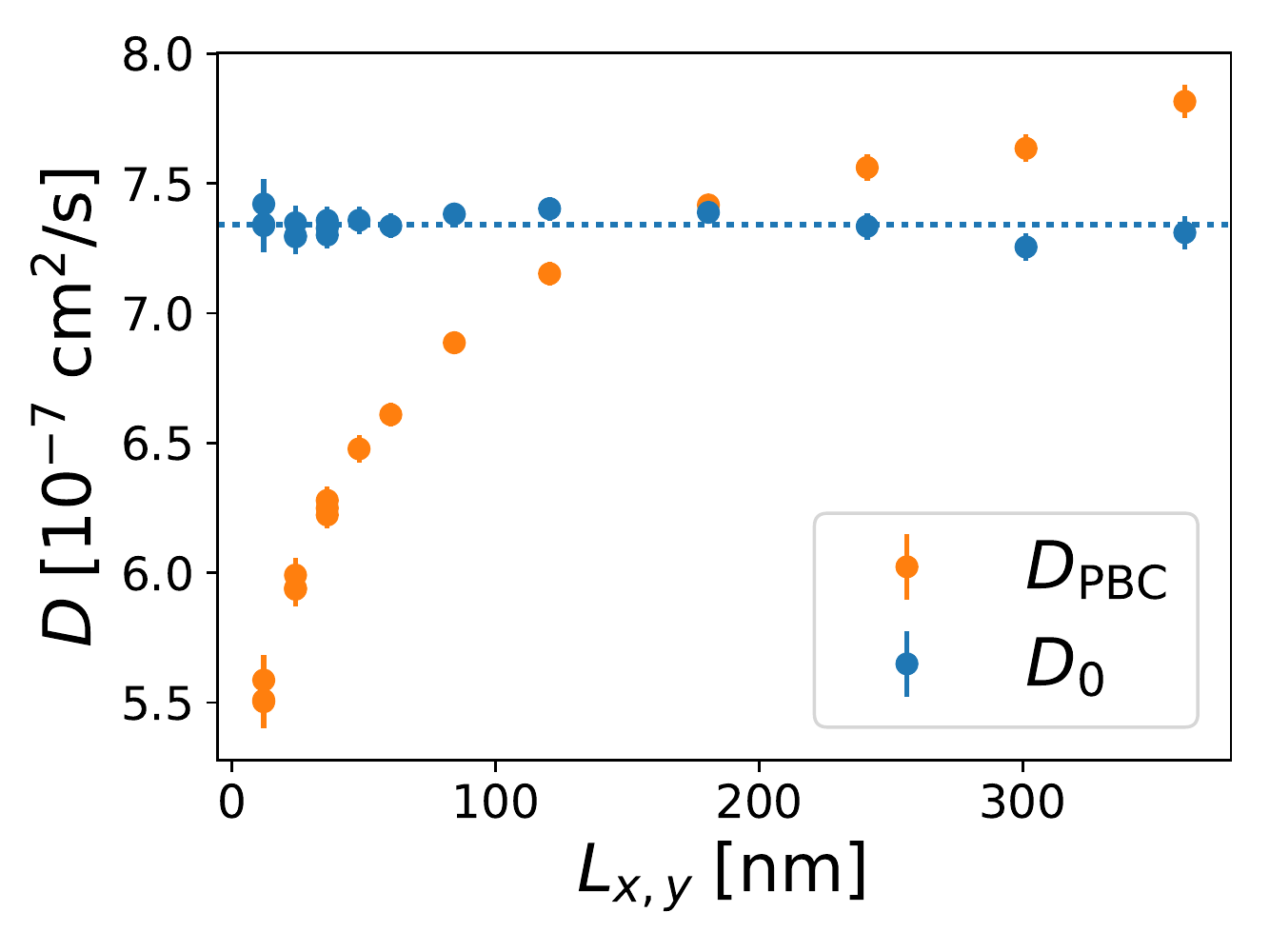}
	\includegraphics[width=0.49\linewidth]{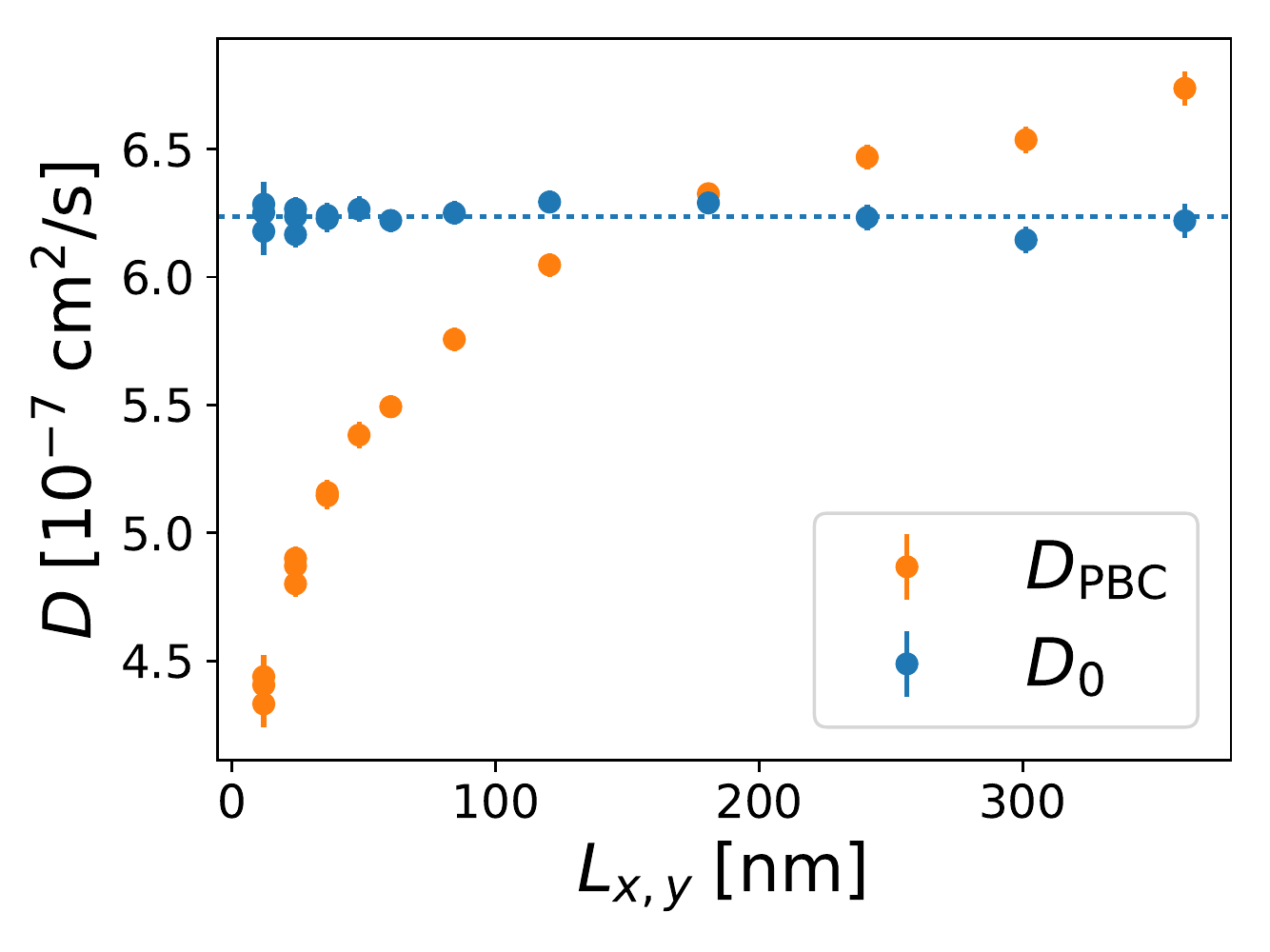}	
	$\qquad$ POPE (upper leaflet) $\qquad\qquad\qquad\qquad\qquad\qquad\;$ POPE (lower leaflet)\\
	\includegraphics[width=0.49\linewidth]{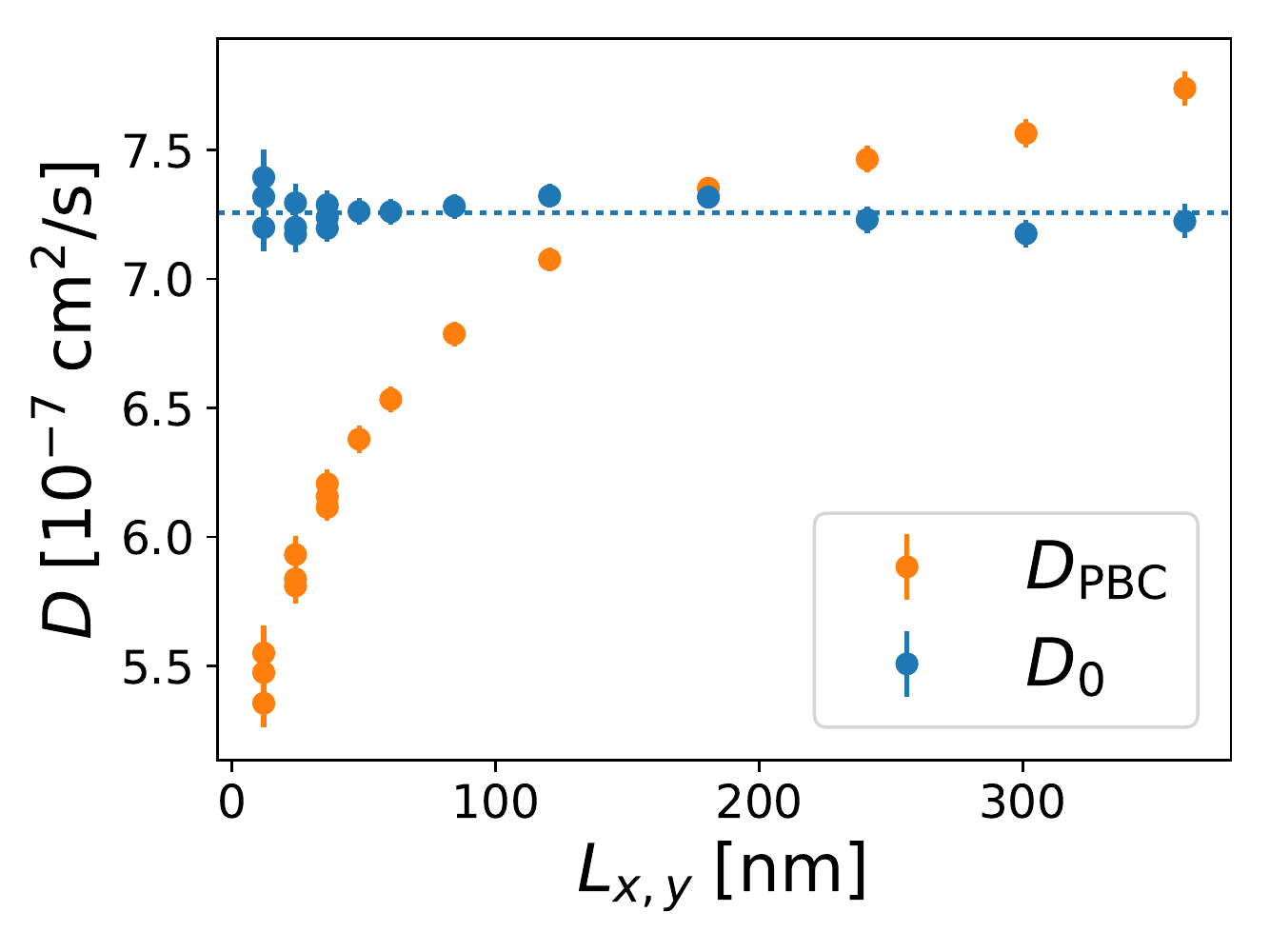}
	\includegraphics[width=0.49\linewidth]{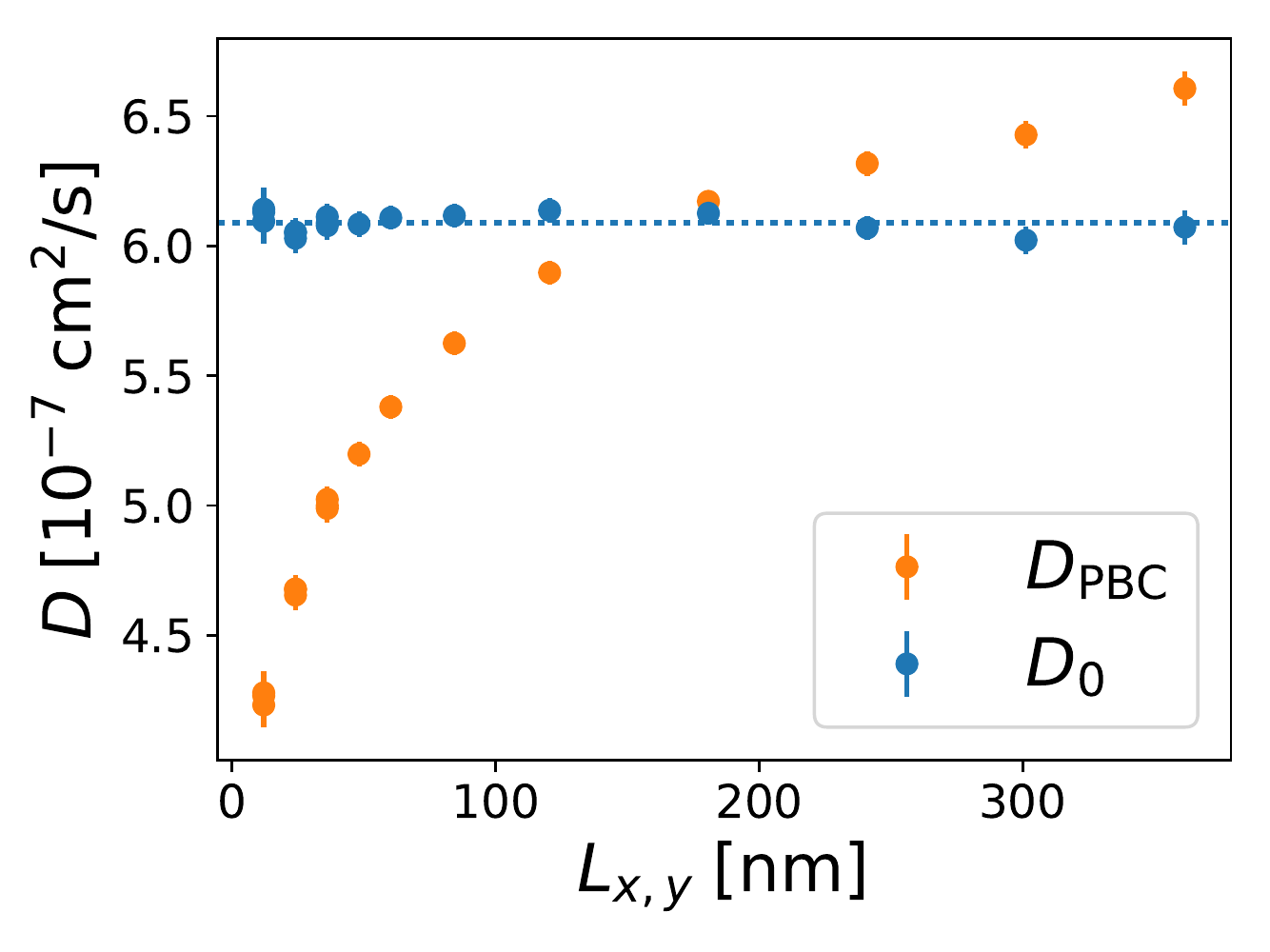}\\
    \caption{Diffusion coefficient of all components in the simulations of ANT1 proteins in a cardiolipin-containing POPC/POPE membrane with the infinite-system value fitted with fixed fluid viscosity $\eta_f = 8.4\times 10^{-4}\,\mathrm{Pa\,s}$.
           \label{fig:study-lb-detailed}}
\end{figure}

\begin{figure}[ht]
	\centering
	$\qquad\qquad$ Protein \hspace{4cm} Cardiolipin (lower leaflet)\\
	\includegraphics[width=0.49\linewidth]{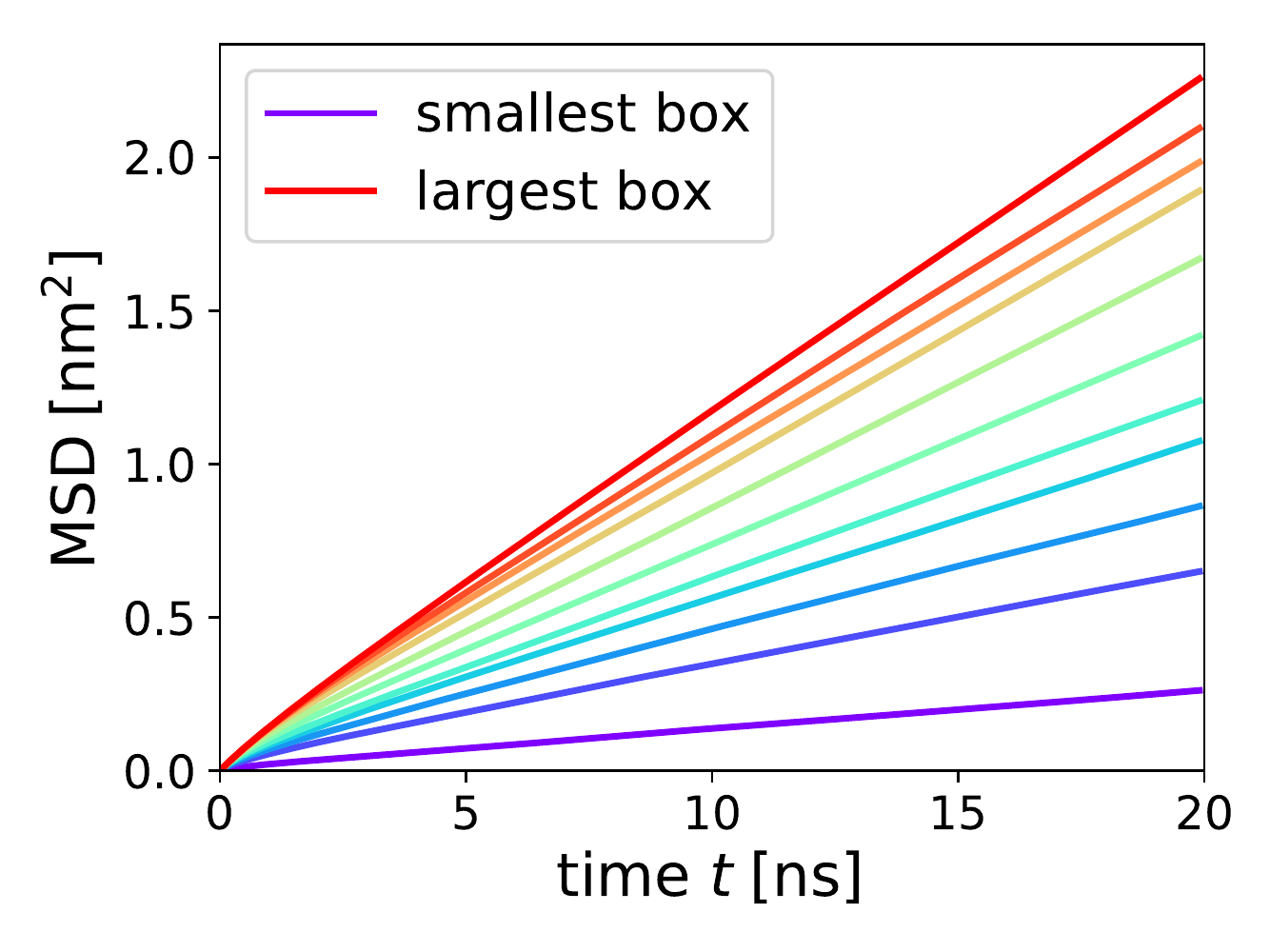}
	\includegraphics[width=0.49\linewidth]{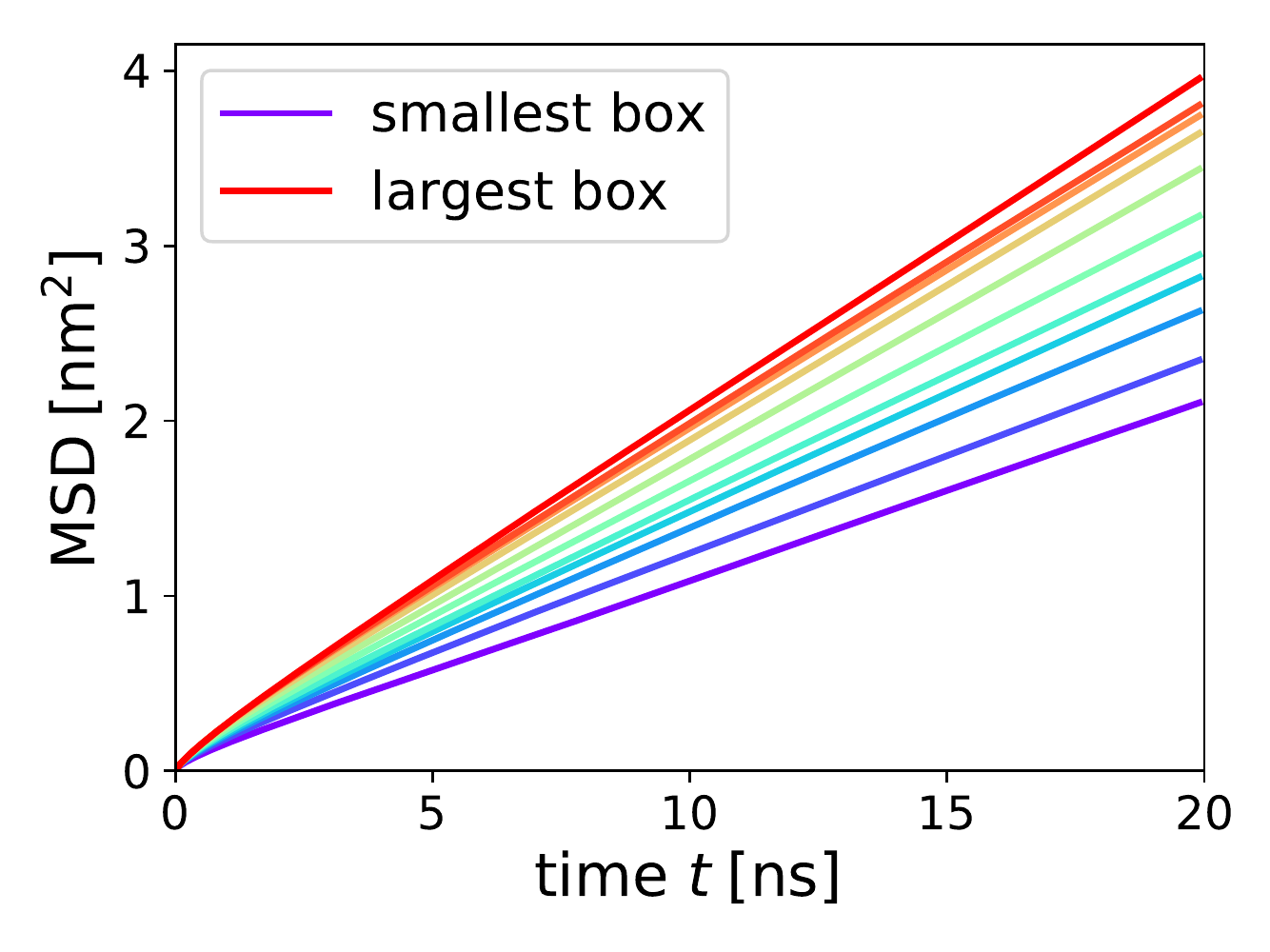}\\
	$\qquad$ POPC (upper leaflet) $\qquad\qquad\qquad\;$ POPC (lower leaflet)\\
	\includegraphics[width=0.49\linewidth]{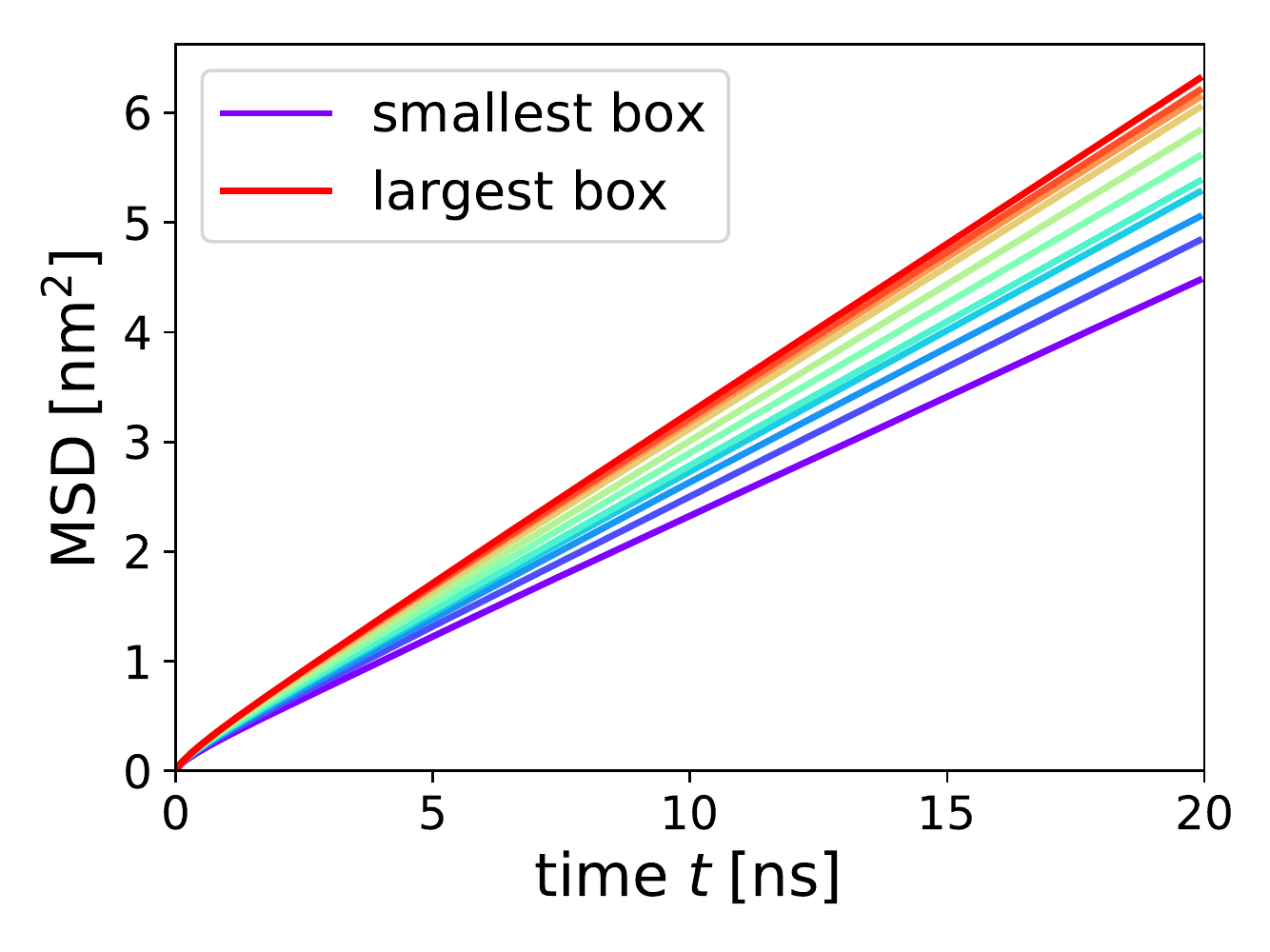}
	\includegraphics[width=0.49\linewidth]{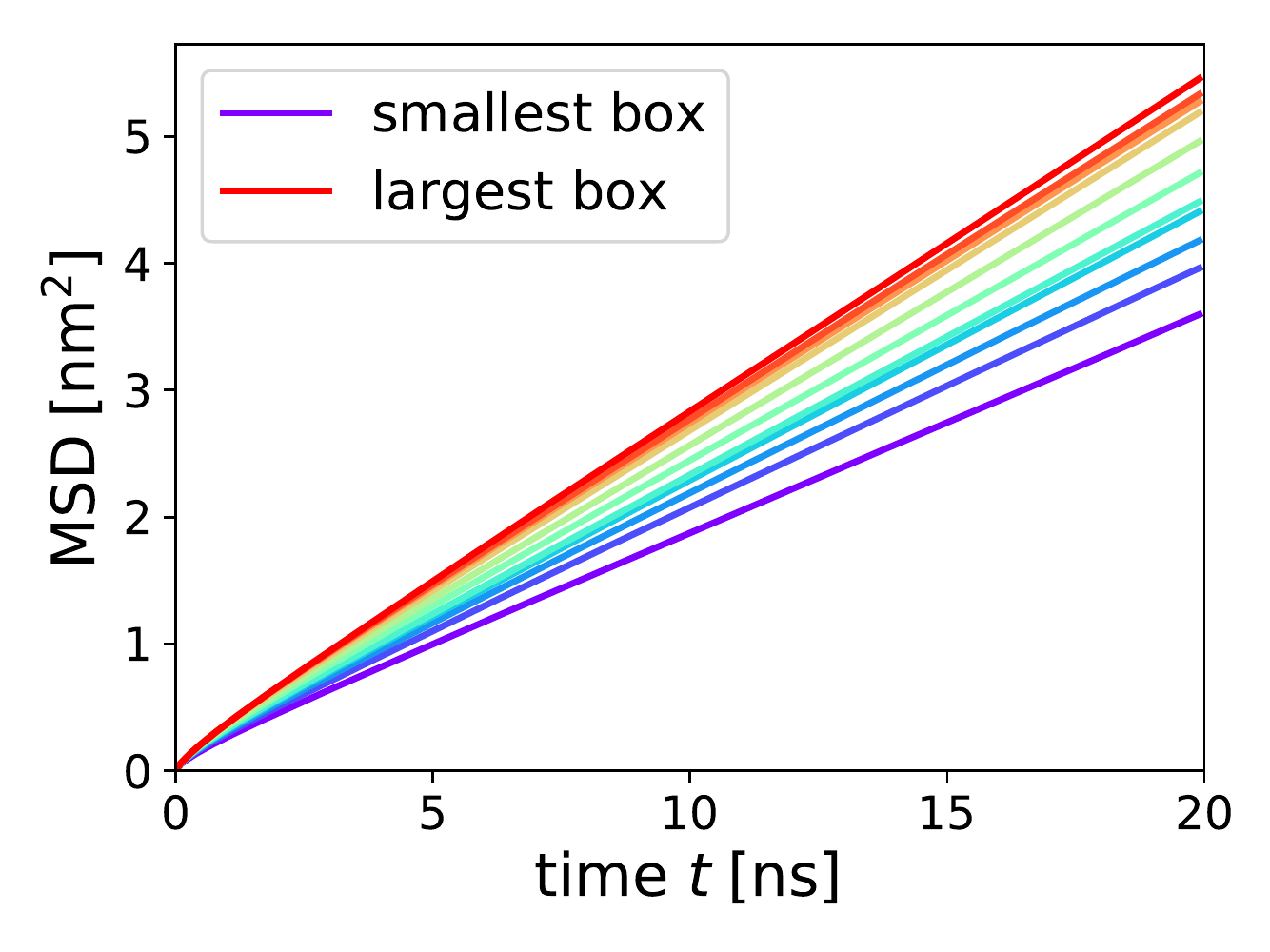}\\	
	$\qquad$ POPE (upper leaflet) $\qquad\qquad\qquad\;$ POPE (lower leaflet)\\
	\includegraphics[width=0.49\linewidth]{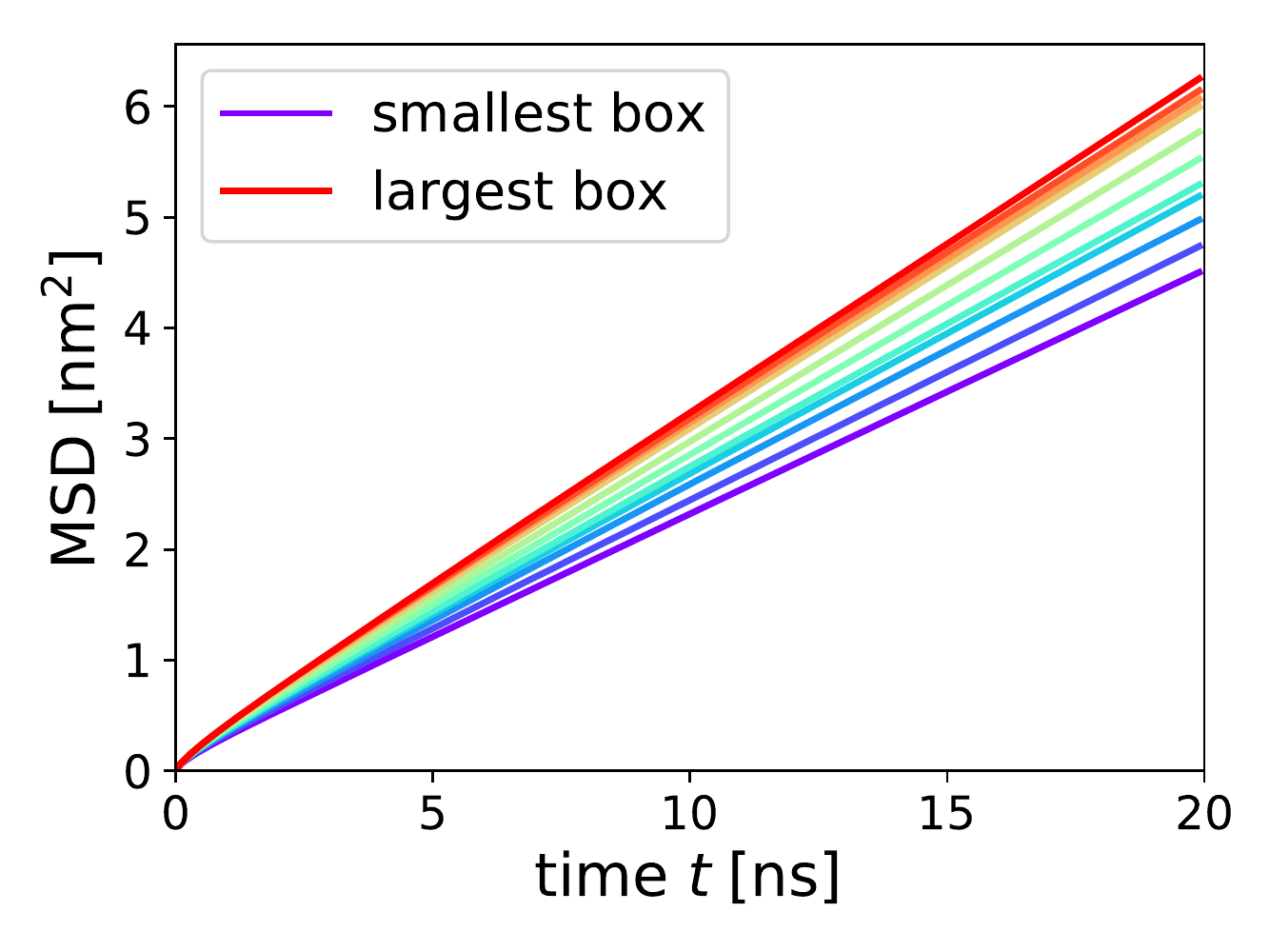}
	\includegraphics[width=0.49\linewidth]{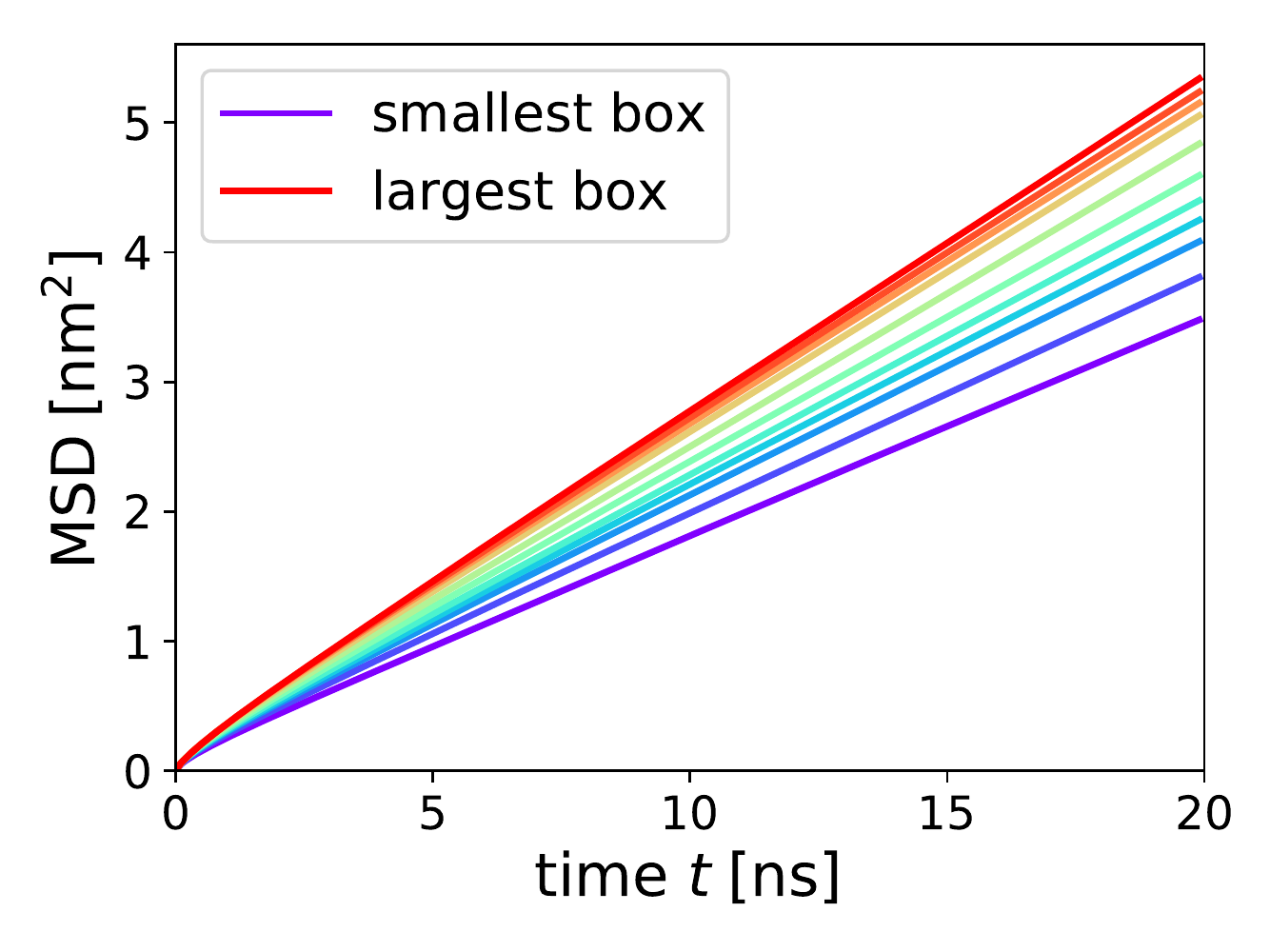}\\
    \caption{MSD curves of all components in the simulations of ANT1 proteins in a cardiolipin-containing POPC/POPE membrane.
           \label{fig:study-lb-msd}}
\end{figure}

\begin{table}[ht]
\begin{center}
\begin{tabular}{c|c|c}
membrane & $\eta_{m}$ & $D_{0}$  \\ 
component & $[10^{-11}\,\mathrm{Pa\,s\,m}]$ & $[10^{-7}\;\mathrm{cm^2/s}]$ \\ \hline
Protein     & 4.08 & 2.15  \\
Cardiolipin & 4.30 & 4.34  \\
POPC, outer & 4.47 & 7.34  \\
POPC, inner & 4.42 & 6.24  \\
POPE, outer & 4.43 & 7.26  \\
POPE, inner & 4.35 & 6.09  \\
\end{tabular}
\end{center}
\caption{Membrane viscosity $\eta_m$ and infinite-system diffusion coefficients $D_0$ from individual fits using the approximation Eq.~(2) for each component. The viscosity obtained from a global fit is $\eta_{m} = 4.36\times 10^{-11}\;\mathrm{Pa\,s\,m}$. The results deviate by typically $\approx$~2~{\%} and maximally by 5~{\%} from those obtained by using the full Oseen correction (see Table~\ref{tab:ant1-fits-oseen}).
}
\end{table}

\begin{table}[ht]
\begin{center}
\begin{tabular}{c|c|c|c|c}
membrane & $\eta_{m}$ (monotopic) & $D_{0}$ (monotopic)  & $\eta_{m}$ (bitopic) & $D_{0}$ (bitopic)  \\ 
component & $[10^{-11}\,\mathrm{Pa\,s\,m}]$ & $[10^{-7}\;\mathrm{cm^2/s}]$ & $[10^{-11}\,\mathrm{Pa\,s\,m}]$ & $[10^{-7}\;\mathrm{cm^2/s}]$ \\ \hline
Protein     & - & - & 4.08 & 2.04 \\
Cardiolipin & 4.45 & 4.26 & 4.30 & 4.24 \\
POPC, outer & 4.61 & 7.26 & 4.47 & 7.24 \\
POPC, inner & 4.57 & 6.16 & 4.41 & 6.13 \\
POPE, outer & 4.57 & 7.17 & 4.43 & 7.16 \\
POPE, inner & 4.50 & 6.01 & 4.34 & 5.99 \\
\end{tabular}
\end{center}
\caption{Membrane viscosity $\eta_m$ and infinite-system diffusion coefficients $D_0$ from individual fits for each component using the Oseen correction Eq.~(1) with the monotopic and the bitopic formalism, respectively. For the monotopic case, $b=2.9\times 10^6$ Pa~s/m was used.
\label{tab:ant1-fits-oseen}
}
\end{table}

\clearpage
\newpage

\subsection{Details of the Plasma Membrane Model}

\begin{table}[h]
\centering
\begin{tabular}{c|c|c|c|c|c}
mult. & num. of lipids & num. of solvent beads & box width $\langle L \rangle$ [nm] & box height $\langle L_z \rangle$ [nm] & sim. time [$\mathrm{\mu s}$] \\ \hline
 1 &  28535 &  292306 &  71.39(6) & 11.30(2) & 2.00 \\
 4 & 114140 & 1169224 & 142.80(6) & 11.29(1) & 2.00 \\
 9 & 256815 & 2630754 & 214.20(2) & 11.29(1) & 1.70 \\
16 & 456560 & 4676896 & 285.59(3) & 11.30(1) & 2.00
\end{tabular}
\caption{System parameters of the plasma membrane simulations.} 
\label{tab:system-parameters-plasmem}
\end{table}

\begin{figure}[ht]
	\centering
	\includegraphics[width=1.0\linewidth]{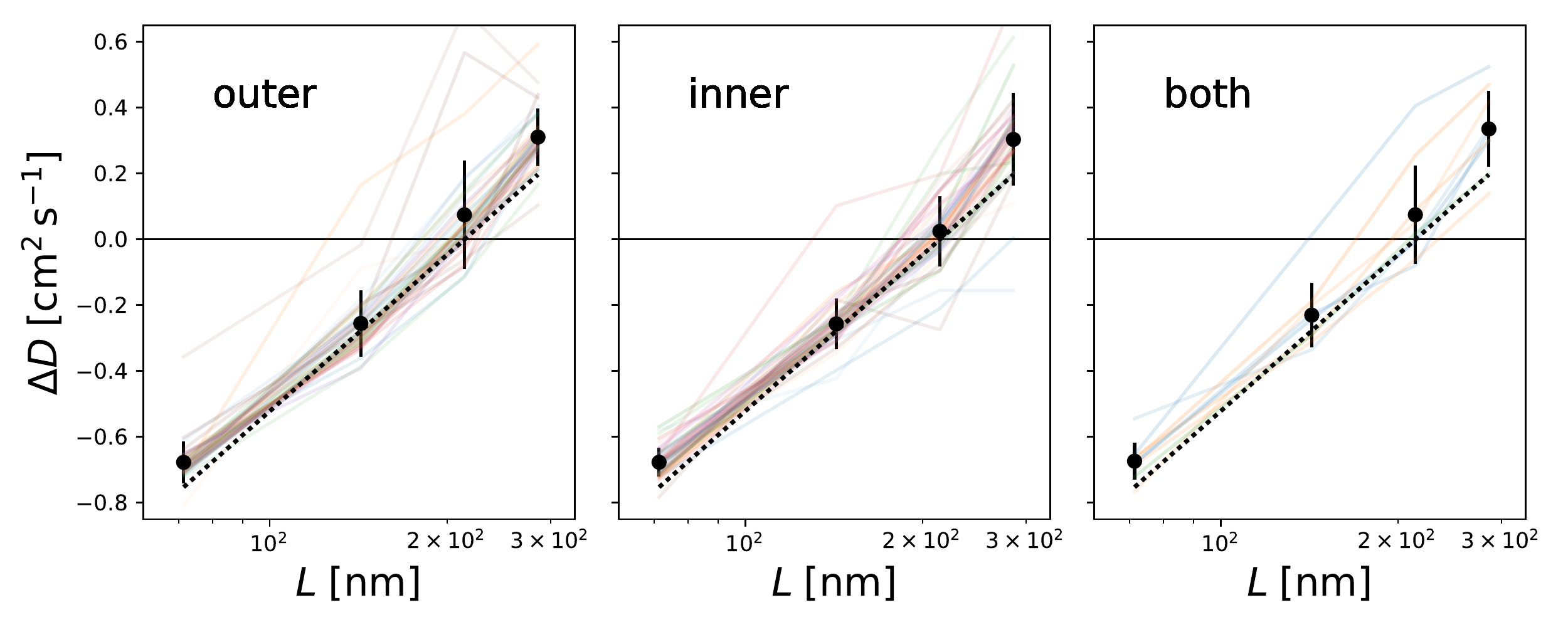}
	\includegraphics[width=1.0\linewidth]{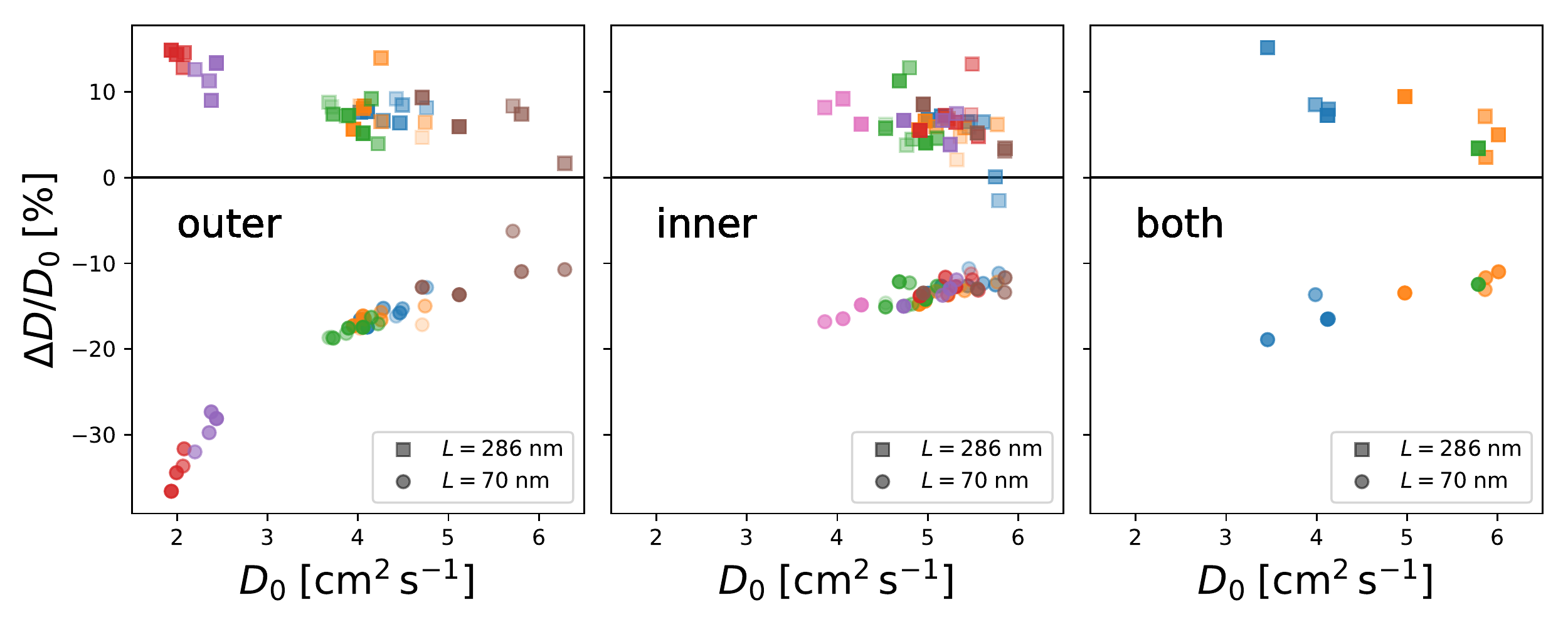}
    \caption{Comparisons of the results for the plasma membrane. Top: Deviations of the diffusion coefficients in the simulations from the respective infinite-system value (shaded colored lines) with their average (black circles) and the theoretical prediction (dashed line). Bottom: Relative errors of the diffusion coefficients of different lipid species depending on their infinite-system diffusion coefficient. Squares show the results of the largest system and circles those of the smallest system under study.
    \label{fig:study-lc-detailed}}
\end{figure}

\begin{figure}[ht]
	\includegraphics[width=0.9\linewidth]{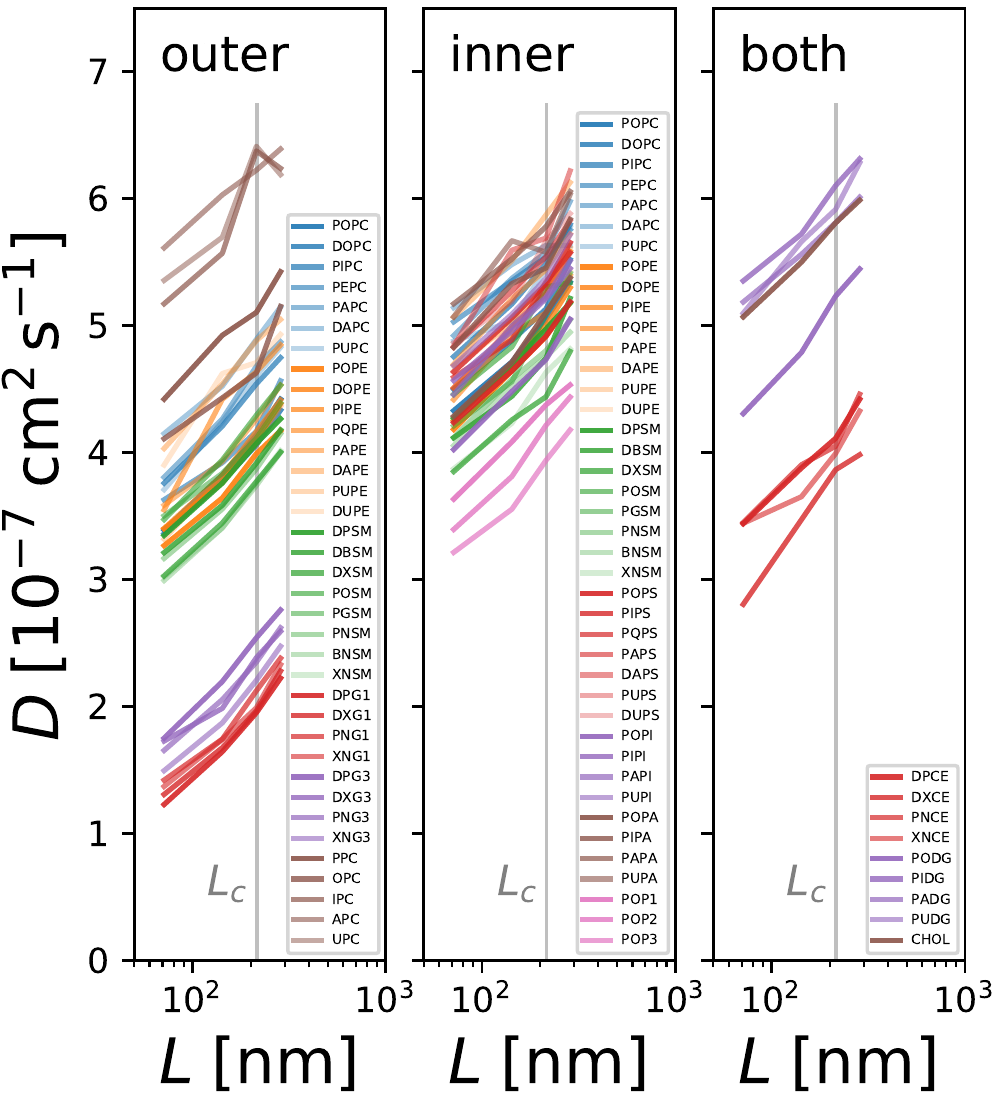}
    \caption{ \label{fig:study-lc-all}
Diffusion coefficients for membrane components that remain in each leaflet (inner/outer) and for those that jump between both leaflets. Colors are assigned according to the lipid types present in each group. Present in the inner as well as in the outer leaflet are phosphatidylcholines (PC - blue), phosphatidylethanolamines (PE - orange), and sphingomyelins (SM - green). Present only in the outer leaflet are monosialotetrahexosylgangliosides (GM1/G1 - red), monosialodihexosylganglioside (GM3/G3 - purple), and lysophosphatidylcholines (LPC/C - brown). Present only in the outer leaflet are phosphatidylserines (PS - red), phosphatidylinositols (PI - purple), phosphatidic acids (PA - brown), and phosphatidylinositol(1-3)phosphates (PIP/P(1-3) - pink). Components that flip between leaflets are ceramides (CER/CE - red), diacylglycerols (DAG/DG - purple) and cholesterol (brown).  The vertical grey lines indicate $L_c$, the box width at which the infinite-system value of the diffusion coefficient is reproduced.}
\end{figure}

\begin{figure}[ht]
	\includegraphics[width=0.75\linewidth]{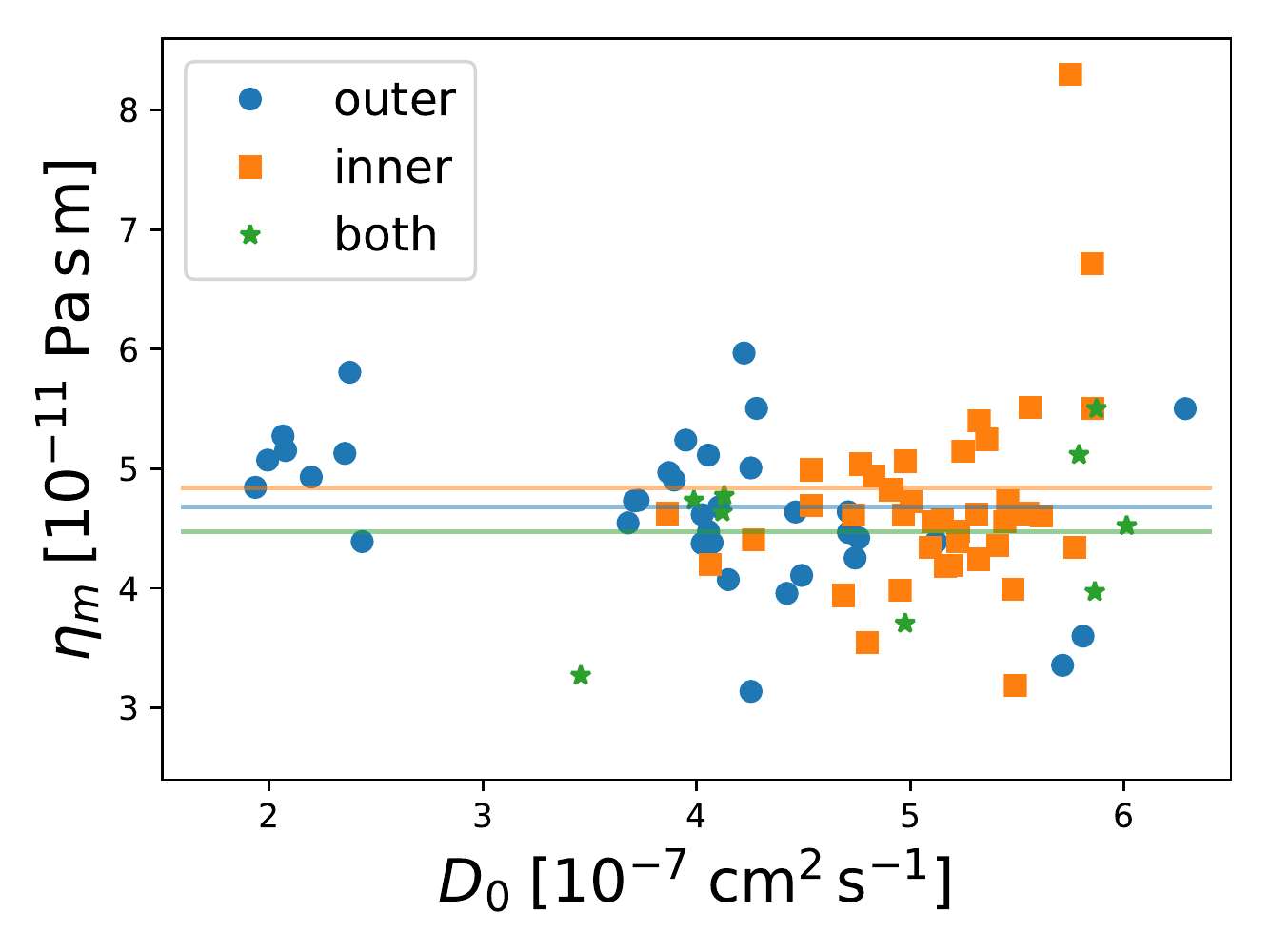}
    \caption{ \label{fig:study-lc-allviscosities}
Membrane viscosity $\eta_m$ obtained from fits to each single component of the model plasma membrane, grouped by leaflet (``both'' indicates components that flip between the inner and outer leaflets).
}
\end{figure}

\begin{table}[ht]
\begin{center}
\begin{tabular}{c|c}
leaflet & $\eta_{m,i}\;[10^{-11}\,\mathrm{Pa\,s\,m}]$  \\ \hline
outer   & $4.68$ \\
inner   & $4.84$ \\
both    & $4.47$ \\ \hline
global  & $4.73$
\end{tabular}
\end{center}
\caption{Membrane viscosity $\eta_m$ from fits for each leaflet and for a global fit.}
\end{table}

\end{document}